\documentclass[a4paper,11pt]{article}

\usepackage{amsfonts,amssymb,amsmath,amsthm,dsfont,xfrac,xspace}
\usepackage{fullpage}
\usepackage{graphicx}
\usepackage{subfig}
\usepackage{float}
\usepackage{cite}
\usepackage{hyperref}
\usepackage{algorithmic}
\graphicspath{{./},{fig/}}
\makeatletter
\let\NAT@parse\undefined
\makeatother
\usepackage[sort&compress, numbers]{natbib}
\usepackage{hyperref}

\newsavebox{\ieeealgbox}
\newenvironment{boxedalgorithmic}
  {\begin{lrbox}{\ieeealgbox}
   \begin{minipage}{\dimexpr\columnwidth-2\fboxsep-2\fboxrule}
   \begin{algorithmic}[1]}
  {\end{algorithmic}
   \end{minipage}
   \end{lrbox}\noindent\fbox{\usebox{\ieeealgbox}}}

\newcommand{\bsy}[1]{\boldsymbol{#1}}
\newcommand{\supp}{\text{supp}}

\DeclareMathOperator*{\argmax}{arg\,max}

\newtheorem{thm}{Theorem}
\newtheorem{lem}{Lemma}[thm]
\renewcommand{\thelem}{\the \numexpr (\value{thm}+1) \relax.\arabic{lem}}

\newtheorem{lemSA}{Lemma}

\newcounter{algoCounter}

\title{On the Noise Robustness of Simultaneous Orthogonal Matching Pursuit} 

\author{ Jean-Fran\c{c}ois Determe\thanks{Jean-Fran\c{c}ois Determe and Fran\c{c}ois Horlin are with the OPERA Wireless Communications Group, Universit\'e libre de Bruxelles, 1050 Brussels, Belgium. E-mail: jdeterme@ulb.ac.be, fhorlin@ulb.ac.be. Jean-Fran\c{c}ois Determe is funded by the Belgian National Science Foundation (F.R.S.-FNRS).}  \quad J\'er\^{o}me Louveaux\footnotemark[2] \quad  Laurent Jacques\thanks{Laurent Jacques, J\'{e}r\^{o}me Louveaux, and Jean-Fran\c{c}ois Determe are with the ICTEAM departement, Universit\'e catholique de Louvain. E-mail: laurent.jacques@uclouvain.be, jerome.louveaux@uclouvain.be. Laurent Jacques is funded by the Belgian National Science Foundation (F.R.S.-FNRS).} \quad Fran\c{c}ois Horlin\footnotemark[1] }

\begin{document}
\maketitle

\begin{minipage}{0.9 \linewidth}
\textbf{\underline{IEEE copyright notice}} -- published paper: J.~Determe, J.~Louveaux, L.~Jacques and F.~Horlin, ``On the Noise Robustness of Simultaneous Orthogonal Matching Pursuit," in \textit{IEEE Transactions on Signal Processing}, vol. 65, no. 4, pp. 864-875, Feb. 15 2017. doi: 10.1109/TSP.2016.2626244. \url{http://ieeexplore.ieee.org/document/7738592/}
\end{minipage}

\begin{abstract}
	In this paper, the joint support recovery of several sparse signals whose supports exhibit similarities is examined. Each sparse signal is acquired using the same noisy linear measurement process, which returns fewer observations than the dimension of the sparse signals. The measurement noise is assumed additive, Gaussian, and admits different variances for each sparse signal that is measured. Using the theory of compressed sensing, the performance of simultaneous orthogonal matching pursuit (SOMP) is analyzed for the envisioned signal model. The cornerstone of this paper is a novel analysis method upper bounding the probability that SOMP recovers at least one incorrect entry of the joint support during a prescribed number of iterations. Furthermore, the probability of SOMP failing is investigated whenever the number of sparse signals being recovered simultaneously increases and tends to infinity. In particular, convincing observations and theoretical results suggest that SOMP committing no mistake in the noiseless case does not guarantee the absence of error in the noisy case whenever the number of acquired sparse signals scales to infinity. Finally, simulation results confirm the validity of the theoretical results.
\end{abstract}

\section{Introduction}
The recovery of sparse signals, \textit{i.e.}, signals exhibiting a low number of non-zero entries, acquired through noisy linear measurements is a central problem in digital signal processing. This task is further involved when the number of measurements is lower than the dimension of the signal to be recovered. Recently, researchers have paid special attention to this class of problems due to the emergence of the \textit{compressed sensing} (CS) field of research \cite{donoho2006compressed}, which aims at providing reliable recovery methods of sparse signals for which the number of measurements is low.\\

Before describing our problem, we wish to introduce key notions. The cardinality of a set $A$ is denoted by $|A|$. The notation $\lbrack n \rbrack$ denotes the set $\lbrace 1, 2, \dots, n\rbrace$ and $x_j$ refers to the $j$-th entry of $\bsy{x}$. The support of any vector $\bsy{x} \in \mathbb{R}^n$ is defined as $\supp (\bsy{x}) := \lbrace j \in \lbrack n \rbrack : x_j \neq 0\rbrace$. We define $s$-sparse signals as vectors whose supports exhibit a cardinality equal to or less than $s$. Loosely speaking, a signal $\bsy{x} \in \mathbb{R}^n$ is said to be sparse whenever its support is significantly smaller than the dimension of its space, \textit{i.e.}, $|\mathrm{supp}(\bsy{x})| \ll n$. Finally, the notation $A \gtrsim B$ means that $\exists \, c > 0 : A \geq c B$.
\subsection{Objective and signal model} \label{subsec:sigModObj}

We focus on a scenario where the objective is recovering the joint support $\mathcal{S}$ of $K$ sparse signals $\bsy{x}_k \in \mathbb{R}^n$ ($1 \leq k \leq K$), \textit{i.e.}, $\mathcal{S} := \supp (\bsy{X}) := \cup_{k \in \lbrack K \rbrack} \supp (\bsy{x}_k)$, observed by means of a common measurement matrix $\bsy{\Phi} \in \mathbb{R}^{m \times n}$ for which $m < n$. The resulting measurement vectors $\bsy{y}_k \in \mathbb{R}^m$ ($1 \leq k \leq K$) gather the measurements of each sparse signal: $\bsy{y}_k = \bsy{\Phi} \bsy{x}_k + \bsy{e}_k$ where $\bsy{e}_k$ is an additive noise term. It is assumed that $\bsy{e}_k$ is distributed as $\mathcal{N}(0, \sigma_k^2 \bsy{I}_{m \times m})$ and that, for $k_1 \neq k_2$, $\bsy{e}_{k_1}$ and $\bsy{e}_{k_2}$ are statistically independent. The vector of the noise standard deviations is denoted by $\bsy{\sigma} := (\sigma_1, \dots, \sigma_K )^{\mathrm{T}}$.\\

For the sake of simplicity,  Equation (\ref{eq:sigModelNoise}) aggregates the $K$ equations $\bsy{y}_k = \bsy{\Phi} \bsy{x}_k + \bsy{e}_k$ into a single relationship:
\begin{equation}\label{eq:sigModelNoise}
\bsy{Y} = \bsy{\Phi} \bsy{X} + \bsy{E}
\end{equation}
where $\bsy{Y} = \big(\bsy{y}_1, \dots, \bsy{y}_K \big) \in \mathbb{R}^{m \times K}$, $\bsy{X} = \big(\bsy{x}_1, \dots, \bsy{x}_K \big) \in \mathbb{R}^{n \times K}$ and $\bsy{E} = \big(\bsy{e}_1, \dots, \bsy{e}_K \big) \in \mathbb{R}^{m \times K}$. Note that the signal models incorporating only one sparse vector are called single measurement vector (SMV) models while those where $K > 1$ sparse signals are measured are referred to as multiple measurement vector (MMV) models \cite{eldar2009robust}.\\

The MMV signal model above applies to several scenarios. For example, in \cite[Section IV.B]{malioutov2005sparse}, the source localization problem is studied when using measurements at different time instants. The signal model describing their problem is equivalent to Equation~(\ref{eq:sigModelNoise}), where each measurement vector in $\bsy{Y}$ corresponds to one time instant. In \cite[Section 3]{baron2009distributed}, the authors describe jointly sparse signal models comparable to ours that typically occur in networks of sensors where a possibly large number of sensing nodes exchange their measurements of the same object to reconstruct it. In this case, each individual measurement vector within $\bsy{Y}$ is generated by one sensor. Other applications are described in the survey \cite{rakotomamonjy2011surveying}.\\

In this paper, the \textit{atoms} should be understood as being the columns of $\bsy{\Phi}$, \textit{i.e.}, $\lbrace \bsy{\phi}_j \rbrace_{j \in \lbrack n \rbrack}$ where $\bsy{\phi}_j$ denotes the $j$-th column of $\bsy{\Phi}$. Although the notion of atom is typically used when dealing with dictionaries, we use it here as it simplifies our discussions. Note that most of the results presented in this paper assume that all the atoms of $\bsy{\Phi}$ have a unit $\ell_2$ norm.

\subsection{Detailed contribution}

Our main contribution is a method to analyze SOMP in a noisy setting. The principal technical contribution is an upper bound on the probability that SOMP identifies incorrect entries, \textit{i.e.}, entries not belonging to $\mathcal{S}$, during a prescribed number of iterations. Using quantities describing the reliability of SOMP in the noiseless case, we then show that the probability of incorrect identification decreases exponentially with the number of sparse signals $K$ under the condition that the amplitudes of the sparse signals are sufficiently high when compared to the noise variances.\\

We also establish that the guarantee that SOMP correctly identifies all the entries of the joint support in the absence of noise is not adequate in noisy scenarios when $K$ tends to $\infty$. The development presented in this paper sheds light on this phenomenon by providing two convincing explanations for its existence.\\

As discussed at the end of the paper, an interesting corollary of our upper bound is the asymptotic exact recovery condition (AERC) for SOMP, \textit{i.e.}, a condition ensuring that the probability of SOMP failing falls to $0$ as $K \rightarrow \infty$. The condition we derive actually also guarantees an arbitrarily small probability of error for a sufficiently high value of $K$, which is a stronger result than the AERC. In particular, our very final result involves four fundamental quantities:
\begin{itemize}
	\item The number of sparse signals $K$, which defines how many independent measurement channels are available for the joint recovery.
	\item The quantity $\mu_X(K) :=  \min_{j \in \mathcal{S}} \frac{1}{K} \sum_{k=1}^{K} | X_{j,k} |$, which sets the minimal averaged amplitudes of the coefficients associated with each atom indexed by the support $\mathcal{S}$.
	\item The quantity $\sigma(K)^2 := (1/K) \sum_{k=1}^K \sigma_k^2$ is the average noise power on all the $K$ sparse signals (or measurement channels).
	\item The quantity $\omega_{\sigma}(K) := (1/\sqrt{K}) \| \bsy{\sigma} \|_1/\| \bsy{\sigma} \|_2 \in \lbrack 1/\sqrt{K}; 1 \rbrack$ quantifies how close to a sparse vector $\bsy{\sigma}$ is. In particular, if $\bsy{\sigma}$ is $1$-sparse, then $\omega_{\sigma}(K) = 1/\sqrt{K}$ while $\omega_{\sigma} = 1$ when its entries are identical. In practice, we prefer working with the upper bound $\omega_{\sigma} = \max_{1 \leq K < \infty} \omega_{\sigma}(K)$ belonging to $( 0; 1 \rbrack$.
\end{itemize}

Given the four quantities above, the minimum signal-to-mean-noise ratio $\mathrm{SNR}_{\mathrm{min}}$ is defined as $\min_{1 \leq K < \infty}  \frac{\mu_X(K)}{\sigma(K)}$. It is shown that the probability of SOMP committing at least one error during its first $s+1$ iterations is no more than $p_{\mathrm{err}}$ if
\begin{equation}\label{eq:KMinValApproxIntro}
K \gtrsim \frac{1}{(\alpha \; \mathrm{SNR}_{\mathrm{min}} - \beta \omega_{\sigma})^2} \left( \log n + s \log \frac{2 e |\mathcal{S}|}{s} - \log p_{\mathrm{err}} \right)
\end{equation}
where the condition is valid only if $\mathrm{SNR}_{\mathrm{min}} > \frac{\beta \omega_{\sigma}}{\alpha}$. In particular, if $\mathrm{SNR}_{\mathrm{min}} \gg \frac{\beta \omega_{\sigma}}{\alpha}$, then $\alpha \mathrm{SNR}_{\mathrm{min}} - \beta \omega_{\sigma} \simeq  \alpha \mathrm{SNR}_{\mathrm{min}}$. As clarified later on, the term $\beta$ is upper bounded by $\sqrt{2/\pi}$ while $\alpha$ quantifies the sensing properties of the matrix $\bsy{\Phi}$ as well as the reliability of SOMP decisions in the noiseless case. Note that the condition $\mathrm{SNR}_{\mathrm{min}} > \frac{\beta \omega_{\sigma}}{\alpha}$ testifies to the impossibility of performing correct decisions unless the SNR is above a certain minimal floor level.\\

Finally, some numerical simulations validate our methodology by showing that  the expression appearing in Equation~(\ref{eq:KMinValApproxIntro}) is accurate up to some numerical adjustments aiming at tightening large theoretical constants.

\subsection{Related work}\label{subsec:relatedWork}

In a SMV setting, full support recovery guarantees for OMP with bounded noise signals as well as with Gaussian noises have been proposed in \cite{cai2011orthogonal}. This work also provides criteria on the stopping criteria to guarantee that OMP terminates after having picked all the correct atoms. This contribution has then been slightly refined in \cite{dan2014robustness} to provide conditions independent of the particular support that is to be recovered. \\

Gribonval \textit{et al.} have investigated the performance of SOMP for a problem resembling ours in \cite{gribonval2008atoms}. Their contribution has been to provide a lower bound on the probability of correct full support recovery when the signal to be estimated is sparse and its non-zero entries are statistically independent mean-zero Gaussian random variables.

\subsection{Outline}

First of all, Section~\ref{sec:SOMP} describes SOMP and related quantities. Technical prerequisites are delivered to the reader in Section~\ref{sec:prerequisites}. In Section~\ref{sec:robustnessNoNoise}, we present some results on SOMP in the noiseless case that will be used afterwards. Then, Section~\ref{sec:genResults} provides upper bounds on the probability that SOMP fails to identify a correct atom at a fixed iteration. These results are finally exploited in Section~\ref{sec:specificResults} to deliver usable and easily interpretable upper bounds on the probability that SOMP includes at least one incorrect entry to the estimated support for a prescribed number of iterations. Numerical results presented in Section~\ref{sec:numRes} confirm the validity of our results. The conclusion then follows. Most of the technicalities are reported in the Appendix to simplify the presentation in the core of the paper. 
\subsection{Conventions}\label{subsec:conventions}
We find useful to introduce the main notations used in this paper. For $1 \leq p < \infty$ and $\bsy{x} \in \mathbb{R}^n$, we have $\| \bsy{x} \|_p := (\sum_{j = 1}^n | x_j |^p)^{1/p}$ and $\| \bsy{x} \|_{\infty} := \max_{j \in \lbrack n \rbrack} |x_j|$. With  $\bsy{\Phi} \in \mathbb{R}^{m \times n}$,  we define $\|\bsy{\Phi} \|_{p \rightarrow q}$ as \cite[Equation A.8]{foucart2013mathematical} $\|\bsy{\Phi} \|_{p \rightarrow q} := \sup_{\|\bsy{\phi} \|_p = 1} \left\|\bsy{\Phi} \bsy{\phi}\right\|_q$ where $1 \leq p, q \leq \infty$. In particular, with $\bsy{A} \in \mathbb{R}^{n \times K}$, $\|\bsy{A} \|_{\infty \rightarrow \infty}$ is equal to $\max_{j \in \lbrack n \rbrack} \sum_{k=1}^K |A_{j,k}|$ \cite[Lemma A.5]{foucart2013mathematical}. Unless otherwise specified, every vector is to be understood as a column vector. Also, for $\mathcal{S} \subseteq \lbrack n \rbrack$, $\bsy{x}_{\mathcal{S}}$ denotes the vector formed by the entries of $\bsy{x}$ indexed within $\mathcal{S}$. In a likewise fashion, for $\bsy{\Phi} \in \mathbb{R}^{m \times n}$, we define $\bsy{\Phi}_{\mathcal{S}}$ as the matrix formed by the columns of $\bsy{\Phi}$ whose indexes belong to $\mathcal{S}$. The notation $\overline{\mathcal{S}}$ refers to the relative complement of $\mathcal{S}$ with respect to $\lbrack n \rbrack$. The Moore-Penrose pseudoinverse of any matrix $\bsy{\Phi}$ is given by $\bsy{\Phi}^{+}$ while its transpose is denoted by $\bsy{\Phi}^\mathrm{T}$. The range of any matrix $\bsy{\Phi}$, \textit{i.e.}, the space spanned by its columns, is denoted by $\mathcal{R}(\bsy{\Phi})$. The inner product of two vectors $\bsy{x}$ and $\bsy{y}$, written as $\langle \bsy{x} , \bsy{y} \rangle$, is given by $\bsy{x}^{\mathrm{T}} \bsy{y}$. The minimum and maximum eigenvalues of a matrix $\bsy{\Phi}$ are denoted by $\lambda_{\mathrm{min}}(\bsy{\Phi})$ and $\lambda_{\mathrm{max}}(\bsy{\Phi})$, respectively. The probability measure is given by $\mathbb{P}$ while the mathematical expectation is denoted by $\mathbb{E}$.

\section{Simultaneous orthogonal matching pursuit} \label{sec:SOMP}

Many algorithms have been proposed to solve the joint support recovery problem associated with Equation (\ref{eq:sigModelNoise}). In a SMV setting, canonical algorithms include $\ell_1$-minimization \cite{donoho2003optimally, candes2006stable}, matching pursuit (MP) \cite{mallat1993matching} and orthogonal matching pursuit (OMP) \cite{pati1993orthogonal, davis1997adaptive}. While algorithms relying on $\ell_1$-minimization are probably among the most reliable algorithms designed for compressed sensing problems, they often exhibit a higher computational complexity than their greedy counterparts, such as MP and OMP \cite{tropp2007signal}. Greedy algorithms are thus more suited to real time applications.  The well-known algorithms described above for SMV problems admit several extensions within the MMV framework. Specifically, one of the most natural generalization of OMP is SOMP \cite{tropp2006algorithms}, which is described in Algorithm~\ref{alg:SOMP}.
\begin{figure}[!h]
	\textsc{Algorithm \refstepcounter{algoCounter}\label{alg:SOMP}\arabic{algoCounter}}:\\ 
	Simultaneous orthogonal matching pursuit (SOMP)\\
	
	\vspace{-2mm}
	\begin{boxedalgorithmic}
		\small
		\REQUIRE $\bsy{Y} \in \mathbb{R}^{m \times K}$, $\bsy{\Phi} \in \mathbb{R}^{m \times n}$, $s \geq 1$
		\STATE Initialization: $\bsy{R}^{(0)} \leftarrow \bsy{Y}$ and $\mathcal{S}_0 \leftarrow \emptyset$
		\STATE $t \leftarrow 0$
		\WHILE{$t < s$}
		\STATE Determine the atom of $\bsy{\Phi}$ to be included in the support: \\ $j_t \leftarrow \mathrm{argmax}_{j \in \lbrack n \rbrack} ( \| (\bsy{R}^{(t)})^{\mathrm{T}} \bsy{\phi}_j \|_1 )$
		\STATE Update the support : $\mathcal{S}_{t+1} \leftarrow \mathcal{S}_{t} \cup \left\lbrace j_t \right\rbrace$
		\STATE Projection of each measurement vector onto $\mathcal{R}(\boldsymbol{\Phi}_{S_{t+1}})$: \\$\bsy{Y}^{(t+1)} \leftarrow \boldsymbol{\Phi}_{\mathcal{S}_{t+1}} \boldsymbol{\Phi}_{\mathcal{S}_{t+1}}^{+} \bsy{Y}$
		\STATE Projection of each measurement vector onto $\mathcal{R}(\boldsymbol{\Phi}_{\mathcal{S}_{t+1}})^{\perp}$~: \\ $\bsy{R}^{(t+1)} \leftarrow \bsy{Y} - \bsy{Y}^{(t+1)}$
		\STATE $t \leftarrow t + 1$
		\ENDWHILE
		\RETURN $\mathcal{S}_s$ \COMMENT{Support at last step}
	\end{boxedalgorithmic}
\end{figure}

As described in Algorithm \ref{alg:SOMP}, SOMP updates the support at each iteration $t$ by including in the current estimated support a single atom $\bsy{\phi}_{j_t}$, whose index is denoted by $j_t$, which maximizes SOMP metric 
$$\| (\bsy{R}^{(t)})^{\mathrm{T}} \bsy{\phi}_j \|_1 = \sum_{k=1}^{K} | \langle \bsy{\phi}_{j},  \bsy{r}^{(t)}_k \rangle |$$
(steps $4$ and $5$). In this description of SOMP, $\bsy{r}_k^{(t)}$ denotes the $k$-th column of the residual matrix $\bsy{R}^{(t)}$ at iteration $t$. During steps $6$ and $7$, each measurement vector $\bsy{y}_k$ is projected onto the orthogonal complement of $\mathcal{R}(\boldsymbol{\Phi}_{\mathcal{S}_{t+1}})$, denoted by $\mathcal{R}(\boldsymbol{\Phi}_{\mathcal{S}_{t+1}})^{\perp}$. In this way, an atom cannot be picked twice since, once included in the support, the projection onto   $\mathcal{R}(\boldsymbol{\Phi}_{\mathcal{S}_{t+1}})^{\perp}$ ensures that $\langle \bsy{\phi}_j, \bsy{r}_k^{(t+1)}\rangle = 0$ if $j \in \mathcal{S}_t$. The algorithm finishes when the prescribed number of iterations, $s+1$, is attained. We now turn to the description of useful quantities related to SOMP.\\%

The atoms indexed by the joint support $\mathcal{S}$ are referred to as the correct atoms while the incorrect atoms are those not indexed by $\mathcal{S}$. The set $\mathcal{P}^{(t)}$ contains the $\binom{|\mathcal{S}|}{t}$ orthogonal projectors $\boldsymbol{\Phi}_{\mathcal{S}_{t}} \boldsymbol{\Phi}_{S_{t}}^{+}$ such that $\mathcal{S}_t \subseteq \mathcal{S}$ and $|\mathcal{S}_t| = t$. $\bsy{P}^{(0)}$ is defined as the zero matrix. Loosely speaking, $\mathcal{P}^{(t)}$ is the set of all the possible orthogonal projectors $\bsy{P}^{(t)}$ at iteration $t$ assuming that only atoms belonging to $\mathcal{S}$ have been picked previously. Enumerating all the possible orthogonal projection matrices at each iteration by means of $\mathcal{P}^{(t)}$ will be necessary later on since not knowing the sequence of the atoms picked by SOMP beforehand requires to consider all the possible projectors. For each iteration $t$, we also define 
\begin{equation}\label{eq:BetaDef}
\beta_{j,k}^{(t, \bsy{P})} := | \langle \bsy{\phi}_j, (\bsy{I} - \bsy{P}) \bsy{\Phi} \bsy{x}_k \rangle |
\end{equation}
where $\bsy{P}$ belongs to $\mathcal{P}^{(t)}$. The value of $\beta_{j,k}^{(t, \bsy{P})}$ consequently is the inner product of the $j$-th atom with the $k$-th column of the residual matrix whenever the orthogonal projector at iteration $t$ is $\bsy{P} \in \mathcal{P}^{(t)}$ and no noise is present.\\

We define $\gamma_{c}^{(t, \bsy{P})}$ as the highest SOMP metric obtained for the correct atoms at iteration $t$ without noise and, similarly, $\gamma_{i}^{(t, \bsy{P})}$ is the counterpart quantity for the incorrect atoms instead:
\begin{equation}\label{eq:defGammacandi}
\gamma_{c}^{(t, \bsy{P})} := \max_{j \in S} \sum_{k=1}^K \beta_{j,k}^{(t, \bsy{P})} \,\; \mathrm{and} \;\, \gamma_{i}^{(t, \bsy{P})} := \max_{j \in \overline{\mathcal{S}}} \sum_{k=1}^K \beta_{j,k}^{(t, \bsy{P})}.
\end{equation}
It is also convenient to define two quantities identifying which are the best correct atom and incorrect atom in the noiseless case, at iteration $t$ and for the orthogonal projection matrix $\bsy{P}$:
\begin{equation*}
j_c^{(t,\bsy{P})} = \argmax_{j \in S} \sum_{k=1}^{K} \beta_{j,k}^{(t, \bsy{P})} \;\; \mathrm{and} \;\; j_i^{(t,\bsy{P})} = \argmax_{j \in \overline{\mathcal{S}}} \sum_{k=1}^{K} \beta_{j,k}^{(t, \bsy{P})}
\end{equation*}
where the subscripts $c$ and $i$ refer to correct and incorrect atoms, respectively.
\section{Technical prerequisites \& Notations}\label{sec:prerequisites}
We now wish to settle key theoretical notions to be used later on for the actual analysis of SOMP.
\subsection{Restricted isometry property and related concepts}\label{subsec:RIPROCDef}
Any $\bsy{\Phi} \in \mathbb{R}^{m \times n}$ exhibits the restricted isometry property (RIP) \cite{candes2006stable} of order $s$ if there exists a constant $\delta_s < 1$ such that
\begin{equation}\label{eq:RIPDef}
(1 - \delta_s) \|\bsy{u}\|_2^2 \leq \|\bsy{\Phi} \bsy{u}\|_2^2 \leq (1 + \delta_s) \|\bsy{u}\|_2^2
\end{equation}
for all $s$-sparse vectors $\bsy{u}$. The smallest $\delta_s$ for which Equation~(\ref{eq:RIPDef}) holds is the Rectricted Isometry Constant (RIC) of order $s$. The RIC of order $s$ is theoretically given by $\delta_s = \max(U_s, L_s)$ where $U_s = \max_{\mathcal{S} \subseteq \left[n \right], |\mathcal{S}| = s} \lambda_{\mathrm{max}}(\bsy{\Phi}_{\mathcal{S}}^{\mathrm{T}} \bsy{\Phi}_{\mathcal{S}}) - 1$ and $L_s = 1 - \min_{S \subseteq \left[n \right], |\mathcal{S}| = s} \lambda_{\mathrm{min}}(\bsy{\Phi}_{\mathcal{S}}^{\mathrm{T}} \bsy{\Phi}_{\mathcal{S}})$. The RIC therefore provides an upper bound on the alteration of the $\ell_2$ norm of sparse vectors after multiplication by $\bsy{\Phi}$. Interestingly enough, $\delta_s < 1$ implies that $\bsy{\Phi}_{\mathcal{S}}$ has full column rank for every support $\mathcal{S}$ of size $|\mathcal{S}| \leq s$ since $\min_{S \subseteq \left[n \right], |\mathcal{S}| = s} \lambda_{\mathrm{min}}(\bsy{\Phi}_{\mathcal{S}}^{\mathrm{T}} \bsy{\Phi}_{\mathcal{S}}) > 0$.
\subsection{Lipschitz functions}\label{subsec:LipschtizDef}
A function $f : \mathbb{R}^K \rightarrow \mathbb{R} : \bsy{g} \mapsto f(\bsy{g})$ is called a Lipschitz function with regard to the $\ell_2$-norm if and only if
\begin{equation}
\exists L > 0 : \forall \bsy{x}, \bsy{y} \in \mathbb{R}^K, \, \left| f(\bsy{x}) - f(\bsy{y}) \right| \leq L \left\| \bsy{x} - \bsy{y} \right\|_2
\end{equation}
holds. The constant $L$ is then referred to as the Lipschitz constant. An interesting property of Lipschitz functions is the following concentration inequality for standard Gaussian random vectors $\bsy{g}$ \cite[Theorem 8.40.]{foucart2013mathematical}:
\begin{equation}
\mathbb{P} \left(f(\bsy{g}) - \mathbb{E}\lbrack f(\bsy{g})\rbrack \geq \varepsilon \right) \leq \exp \left( \dfrac{-\varepsilon^2}{2 L^2}\right).
\end{equation}
This result shows that, for $\bsy{g} \sim \mathcal{N}(0, \bsy{I}_{K \times K})$, $f(\bsy{g})$ concentrates around its expectation and the concentration improves as $L$ decreases. The similar inequality $\mathbb{P} (- f(\bsy{g}) + \mathbb{E}\lbrack f(\bsy{g})\rbrack \geq \varepsilon ) \leq \exp ( -\varepsilon^2/(2 L^2))$, where $\varepsilon > 0$, results from the observation that, if $f$ is Lipschitz, then so is $-f$.

\subsection{On the folded normal distribution}\label{subsec:foldedNormVar}

A recurring distribution in this paper is the folded normal distribution, which is the absolute value of a normal random variable. If $X \sim \mathcal{N}(\beta, \sigma^2)$, then $Y = |X|$ is the associated folded normal random variable. We often refer to $X$ as the underlying normal variable. Similarly, the underlying mean and variances are those of the underlying normal variable. Note that the expectation of the folded random variable is always higher than that of its underlying normal random variable because of the folding of the probability density function (PDF) occurring for values lower than $0$.

\section{Results on SOMP without noise}\label{sec:robustnessNoNoise}
As discussed later on, the reliability of SOMP in the noiseless case determines which noise levels are unlikely to make SOMP detect incorrect atoms. In this section, we thus provide a lower bound on the measure of the noiseless reliability given by $\gamma_{c}^{(t, \bsy{P})} - \gamma_{i}^{(t, \bsy{P})}$, \textit{i.e.}, the quantity that fixes the minimum gap of the SOMP metrics for correct and incorrect atoms in the noiseless case. To do so, we assume the existence of three quantities. The first quantity, denoted by $\Gamma$, quantifies the minimum relative reliability of SOMP decisions:
\begin{equation}\label{eq:DefGamma}
\frac{\gamma_{c}^{(t, \bsy{P})}}{\gamma_{i}^{(t, \bsy{P})}} \geq \Gamma > 1 \; \textrm{for all } t \in \big[0, s \big] \; \textrm{and all} \; \bsy{P} \in \mathcal{P}^{(t)}.
\end{equation}

The remaining quantities $\psi$ and $\tau_{X}$ are entwined and provide a lower bound on the SOMP metric for the best correct atom:
\begin{equation}\label{eq:DefTauX}
\gamma_{c}^{(t, \bsy{P})} \geq \psi  \tau_X \; \textrm{for all } t \in \big[0, s \big] \; \textrm{and all} \; \bsy{P} \in \mathcal{P}^{(t)}.
\end{equation}
The quantity $\psi = \psi(\delta_{|\mathcal{S}|}, |\mathcal{S}|)$ typically  increases as $\delta_{|\mathcal{S}|}$ and $|\mathcal{S}|$ dwindle. This property conveys the idea that a small support size $|\mathcal{S}|$ and a measurement matrix $\bsy{\Phi}$ endowed with good CS properties tend to increase the value of $\gamma_{c}^{(t, \bsy{P})}$. Finally, $\tau_{X}$ is related to the intrinsic energy of the sparse signals $\bsy{x}_k$ prior to their projection by $\bsy{\Phi}$. Therefore, $\tau_{X}$ only depends on $\bsy{X}$. Combining $\Gamma$, $\psi$, and $\tau_X$ yields the desired lower bound on the absolute reliability of SOMP in the noiseless case:
\begin{lemSA}\label{lem:noislessIneqDecisions}
	For every $t \in \big[0, s \big]$ and $\bsy{P} \in \mathcal{P}^{(t)}$, we have
	\begin{equation*}
	\gamma_{c}^{(t, \bsy{P})}  - \gamma_{i}^{(t, \bsy{P})} \geq \left(1 - \dfrac{1}{\Gamma} \right) \psi \tau_X.
	\end{equation*}
	where $\Gamma$, $\psi$, and $\tau_X$ are defined in Equations~(\ref{eq:DefGamma})-(\ref{eq:DefTauX}).
\end{lemSA}
The following subsections aim at providing evidence for the existence of $\Gamma$, $\psi$, and $\tau_X$.  In particular, valid expressions are provided for these parameters in light of previous works in the literature.
\subsection{A lower bound on SOMP relative reliability}\label{subsec:lowBoundGSR}
In this section, we briefly discuss the existence of lower bounds for $\gamma_{c}^{(t, \bsy{P})}/\gamma_{i}^{(t, \bsy{P})}$, \textit{i.e.}, possible expressions for $\Gamma$, quantifying how reliably SOMP distinguishes correct and incorrect atoms in the noiseless case. In \cite{determe2016exact}, we have proposed several expressions of $\Gamma$ that use the RIC and sometimes the restricted orthogonality constant (ROC), thereby only requiring to know of the residual support size or, equivalently, the current iteration number $t$. For example, we have shown that $\frac{\gamma_{c}^{(t, \bsy{P})}}{\gamma_{i}^{(t, \bsy{P})}} \geq \frac{(1 - \delta_{|\mathcal{S}|}) \sqrt{|\mathcal{S}|-1}}{\delta_{|\mathcal{S}|} |\mathcal{S}|}$ for all $t < |\mathcal{S}|$ and for $\delta_{|\mathcal{S}|} < 1$. Similarly, a bound using the support itself can be obtained by a straightforward adaptation of the proof of \cite[Theorem 4.5]{chen2006theoretical}, \textit{i.e.}, $\frac{\gamma_{c}^{(t, \bsy{P})}}{\gamma_{i}^{(t, \bsy{P})}} \geq \frac{1}{\left\|\bsy{\Phi}_{\mathcal{S}}^+ \bsy{\Phi}_{\overline{\mathcal{S}}} \right\|_{1 \rightarrow 1} }$ if $\bsy{\Phi}_{\mathcal{S}}$ has full column rank. Note that in both papers, the bounds hold only if correct decisions have been made during all the iterations preceding $t$. We will not get further into the details of such matters as our only objective here is to provide evidence for the existence of $\Gamma$.
\subsection{A lower bound on $\gamma_{c}^{(t, \bsy{P})}$} \label{subsec:LBGammac}
We now present a convenient lower bound for $\gamma_{c}^{(t, \bsy{P})}$ and the associated expressions for $\psi$ and $\tau_X$.
\begin{lemSA}\label{lem:lowBNoiselessMetricCorrectAtoms}
	Adapted from \cite{determe2016improving}. If $\bsy{\Phi}$ satisfies the RIP with $|\mathcal{S}|$-th RIC $\delta_{|\mathcal{S}|}  < 1$, then
	\begin{equation*}
	\gamma_{c}^{(t, \bsy{P})} \geq  \dfrac{(1-\delta_{|\mathcal{S}|})(1+\delta_{|\mathcal{S}|})}{1+ \sqrt{|\mathcal{S}|} \delta_{|\mathcal{S}|}}  \min_{j \in \mathcal{S}} \sum_{k=1}^{K} | X_{j,k}|
	\end{equation*}
	where  $\gamma_{c}^{(t, \bsy{P})}$ is defined in Section~\ref{sec:SOMP}.
\end{lemSA}
This lemma shows that $\gamma_{c}^{(t, \bsy{P})}$ depends on the CS properties of $\bsy{\Phi}$ through the RIC $\delta_{|\mathcal{S}|}$ while $ \min_{j \in \mathcal{S}} \sum_{k=1}^{K} | X_{j,k} |$ indicates that the sum of the absolute coefficients of $\bsy{X}$ associated with each atom also influences $\gamma_{c}^{(t, \bsy{P})}$. Lemma~\ref{lem:lowBNoiselessMetricCorrectAtoms} thus provides  $\tau_X = \min_{j \in \mathcal{S}} \sum_{k=1}^{K} | X_{j,k}|$ and $\psi = (1-\delta_{|\mathcal{S}|})(1+\delta_{|\mathcal{S}|})/(1+ \sqrt{|\mathcal{S}|} \delta_{|\mathcal{S}|})$.\\

Notice that, without restrictions on the sparse signal matrix $\bsy{X}$, nothing can be said about the non-maximum SOMP metrics for the atoms belonging to the correct support $\mathcal{S}$, \textit{i.e.}, it might happen that all the SOMP metrics for the correct atoms are zero except for the highest one among them. As a result, we only focus on the correct atom index by $j_c^{(t,\bsy{P})}$ and the associated noiseless SOMP metric $\gamma_{c}^{(t, \bsy{P})}$. A short example is available in Section~\ref{subsec:LBNoiselessMetricCorrectAtomsExample} to prove this statement.

\section{Upper bounds on the probability of SOMP failing at iteration $t$}\label{sec:genResults}

This section provides an upper bound on the probability that SOMP picks an incorrect atom at iteration $t$ given a fixed orthogonal projector $\bsy{P} \in \mathcal{P}^{(t)}$. This time, the derived results include the noise. First of all, we examine the statistical distribution of SOMP metric for a single atom in Section~\ref{subsec:distL1Norm}. The desired upper bound will then be derived.
\subsection{On the distribution of $\| (\bsy{R}^{(t)})^{\mathrm{T}} \bsy{\phi}_j \|_1$} \label{subsec:distL1Norm}
The quantity $\| (\bsy{R}^{(t)})^{\mathrm{T}} \bsy{\phi}_j \|_1 = \sum_{k=1}^{K} | \langle \bsy{\phi}_{j},  \bsy{r}^{(t)}_k \rangle |$ ultimately defines which atom is picked at each iteration. It is therefore interesting to determine its statistical distribution. We have
\begin{align*}
\sum_{k=1}^{K} | \langle \bsy{\phi}_{j},  \bsy{r}^{(t)}_k \rangle | & = \sum_{k=1}^{K} | \langle \bsy{\phi}_{j},  (\bsy{I} - \bsy{P}^{(t)}) (\bsy{\Phi} \bsy{x}_k + \bsy{e}_k) \rangle | \\
& = \sum_{k=1}^{K} | \langle \bsy{\phi}_j^{(t)},  \bsy{\Phi} \bsy{x}_k \rangle + \langle \bsy{\phi}_j^{(t)},  \bsy{e}_k \rangle |
\end{align*} where $\bsy{\phi}_j^{(t)} := (\bsy{I} - \bsy{P}^{(t)}) \bsy{\phi}_{j}$. Let us consider a fixed projection matrix $\bsy{P}^{(t)} = \bsy{P}$. It is easy to prove that $\left\langle (\bsy{I} - \bsy{P})\bsy{\phi}_j,  \bsy{e}_k \right\rangle$ is distributed as $\mathcal{N}(0, (\sigma_{j, k}^{(\bsy{P})})^2)$ where 
\begin{equation}
\sigma_{j, k}^{(\bsy{P})} := \| (\bsy{I} - \bsy{P}) \bsy{\phi}_{j} \|_2 \sigma_k  \leq \sigma_k 
\end{equation}
provided that $\| \bsy{\phi}_j \|_2 = 1$ .We define the related noise standard deviation vectors 
\begin{equation}
\bsy{\sigma}_j^{(\bsy{P})} := \| (\bsy{I} - \bsy{P}) \bsy{\phi}_{j} \|_2 \bsy{\sigma}
\end{equation}
for each atom. It is thereby possible to replace the term $\langle \bsy{\phi}_j^{(t)},  \bsy{e}_k \rangle$ by $\sigma_{j, k}^{(\bsy{P})} g_k$ where $g_k \sim \mathcal{N}(0, 1)$. Moreover,  $| \langle \bsy{\phi}_j^{(t)},  \bsy{\Phi} \bsy{x}_k \rangle + \sigma_{j, k}^{(\bsy{P})} g_k |$ is distributed as $| |\langle \bsy{\phi}_j^{(t)},  \bsy{\Phi} \bsy{x}_k \rangle | + \sigma_{j, k}^{(\bsy{P})} g_k |$. Hence, $\| (\bsy{R}^{(t)})^{\mathrm{T}} \bsy{\phi}_j \|_1$ is a sum of $K$ folded normal random variables and, since $| \langle \bsy{\phi}_j^{(t)},  \bsy{\Phi} \bsy{x}_k \rangle | = \beta_{j,k}^{(t, \bsy{P})}$,  it admits the same distribution as:
\begin{equation}\label{eq:deffjtP}
f_{j}^{(t, \bsy{P})} : \mathbb{R}^K \rightarrow \mathbb{R} : \bsy{g} \mapsto \sum_{k=1}^K \left| \beta_{j,k}^{(t, \bsy{P})} + \sigma_{j, k}^{(\bsy{P})} g_k \right|.
\end{equation}
\subsection{On the probability of SOMP picking an incorrect atom at iteration $t$}\label{subsec:cdntCorrectDecision}

In this section, we provide a general upper bound on the probability that SOMP picks an incorrect atom at iteration $t$ and for a fixed orthogonal projector $\bsy{P} \in \mathcal{P}^{(t)}$. The idea of the proof is to notice that if the metrics associated with every incorrect atom is lower than a real positive number $\alpha$ and the metric associated with one of the correct atom is higher than $\alpha$, then a correct decision will necessarily be made. This approach is pessimistic in the sense that more than one specific correct atoms could be picked. Note that, among all the correct atoms, only the best atom in the noiseless case is considered, \textit{i.e.}, the atom indexed by $j_c^{(t,\bsy{P})}$. The proof of Theorem~\ref{thm:basicProbIneq} is available in the Appendix.
\begin{thm}\label{thm:basicProbIneq}
	For a fixed iteration $t$, let $\bsy{P} \in \mathcal{P}^{(t)}$ and $\bsy{R} = (\bsy{I} - \bsy{P}) \bsy{Y}$, \textit{i.e.}, $\bsy{R}$ is one of the residuals that could be generated by SOMP on the basis of $\bsy{Y}$ at iteration $t-1$ assuming that only correct atoms have been identified. For $\bsy{g} \sim \mathcal{N}(0, \bsy{I}_{K \times K})$ and for all $\alpha > 0$, the probability of SOMP picking an incorrect atom when running one iteration on $\bsy{R}$ is upper bounded by
	\begin{equation}
	\mathbb{P}\left[ f_{j_c^{(t,\bsy{P})}}^{(t, \bsy{P})} (\bsy{g}) \leq \alpha \right] +  \sum_{j \in \overline{\mathcal{S}}} \mathbb{P} \left[ f_{j}^{(t, \bsy{P})} (\bsy{g}) \geq \alpha \right].
	\end{equation}
\end{thm}
In Theorem~\ref{thm:basicProbIneq}, we do not assume that $\bsy{R}$ has been generated on the basis of past iterations of SOMP, \textit{i.e.}, the upper bound we derive is independent of the way $\bsy{R}$ has been obtained. Ignoring this precaution would imply that we were able to upper bound the conditional probability given the event that SOMP succeeded during the previous iterations, which would be more involved than what Theorem~\ref{thm:basicProbIneq} establishes.\\

\begin{figure}[!h]
	\centering
	\hspace*{-3mm}
	\includegraphics[scale=0.8]{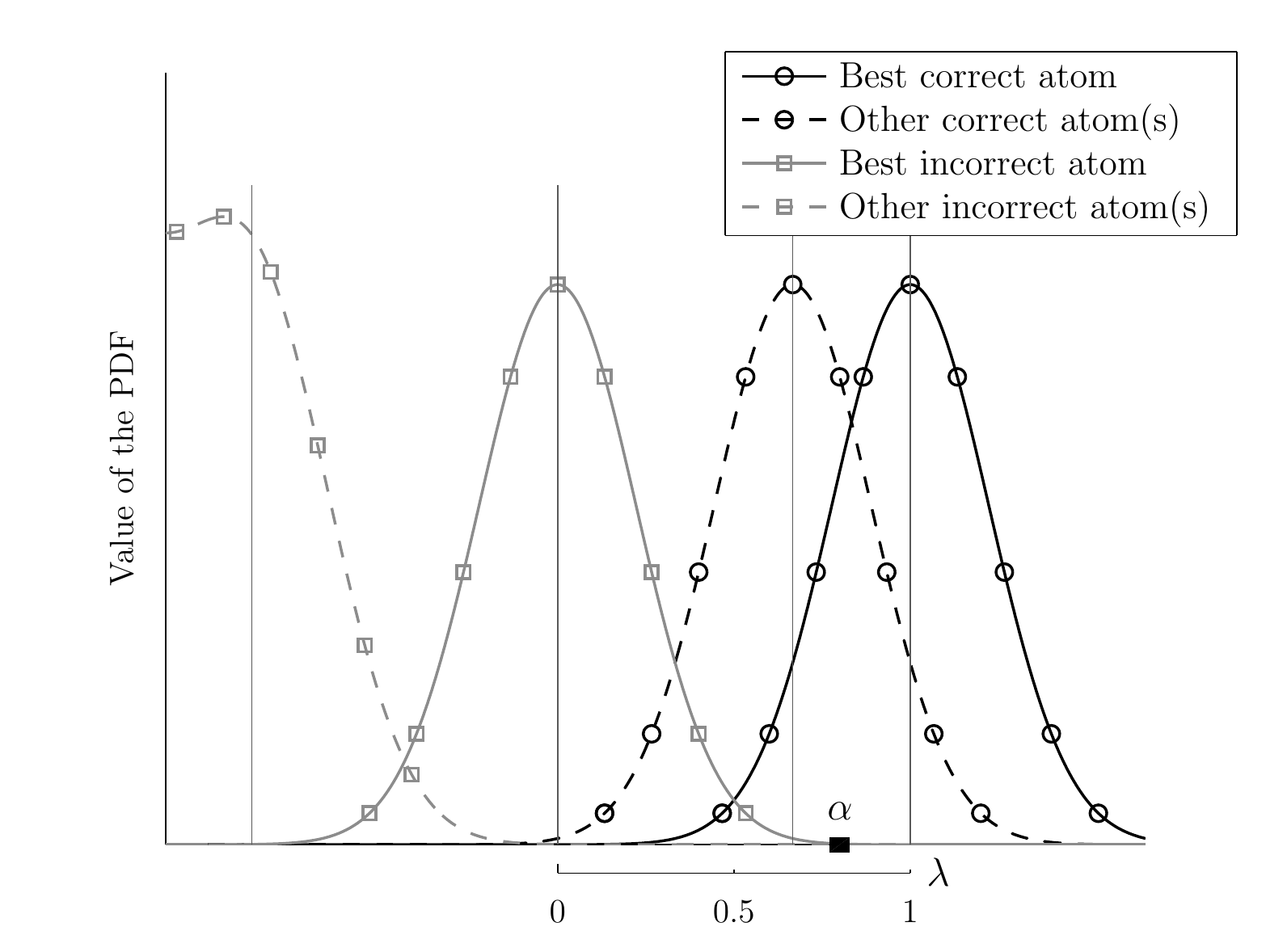}
	\caption{Explanation of Theorem~\ref{thm:concentratedProbIneq} -- Probability density function of $f_{j}^{(t, \bsy{P})} (\bsy{g})$ for $\bsy{g} \sim \mathcal{N}(0, \bsy{I}_{K \times K})$ -- $\bsy{\sigma}_{j_1}^{(\bsy{P})} = \bsy{\sigma}_{j_2}^{(\bsy{P})}$ for $j_1, j_2 \in \lbrack n \rbrack$ -- The vertical lines represent the mean of the folded normal distribution associated with each atom, \textit{i.e.} $\mathbb{E}\big[ f_{j}^{(t, \bsy{P})} (\bsy{g}) \big]$ for the $j$-th atom -- The position of $\alpha$ is obtained for $\lambda = 0.8$ -- Only two atoms belonging to $\mathcal{S}$ (including the best one) and two atoms belonging to $\overline{\mathcal{S}}$ (including the best one) are represented for the sake of clarity.}
	\label{fig:TheoremInterpretation}
\end{figure}

Now that only probabilities of the form $\mathbb{P} \big[ f_{j}^{(t, \bsy{P})} (\bsy{g}) \geq \alpha \big]$ and $\mathbb{P} \big[ f_{j}^{(t, \bsy{P})} (\bsy{g}) \leq \alpha \big]$  intervene, it is appropriate to find upper bounds for these probabilities and use them to produce a more easily interpretable result, \textit{i.e.}, Theorem~\ref{thm:concentratedProbIneq}. The idea of the proof of Theorem~\ref{thm:concentratedProbIneq} is to set the convex combination 
\begin{equation}
\alpha = \lambda \mathbb{E}\big[ f_{j_c^{(t,\bsy{P})}}^{(t, \bsy{P})} (\bsy{g}) \big] + (1-\lambda) \max_{j \in \overline{\mathcal{S}}} \mathbb{E}\lbrack f_{j}^{(t, \bsy{P})} (\bsy{g}) \rbrack,
\end{equation}
\textit{i.e.}, express $\alpha$ in Theorem~\ref{thm:basicProbIneq} relatively to $\mathbb{E}\big[ f_{j_c^{(t,\bsy{P})}}^{(t, \bsy{P})} (\bsy{g}) \big]$ and $\max_{j \in \overline{\mathcal{S}}} \mathbb{E}\lbrack f_{j}^{(t, \bsy{P})} (\bsy{g}) \rbrack$, and then use the concentration inequalities in Section~\ref{subsec:LipschtizDef} in conjunction with the fact that $f_{j}^{(t, \bsy{P})} (\bsy{g})$ is Lipschitz (see Lemma~\ref{lem:fisLipschitz}). It has been chosen to set $\lambda = 0.5$ to simplify the final result. A visual interpretation of Theorem~\ref{thm:concentratedProbIneq} is depicted in Figure~\ref{fig:TheoremInterpretation}. For the sake of simplicity, the figure considers identical noise levels for each atom (remember that $\bsy{\sigma}_j^{(\bsy{P})} := \| (\bsy{I} - \bsy{P}) \bsy{\phi}_{j} \|_2 \bsy{\sigma}$ so that the noise may exhibit different powers for each atom in the general case). The full proofs are available in the Appendix.
\begin{lem}\label{lem:fisLipschitz}
	Each function $f_{j}^{(t, \bsy{P})}$ defined in Equation~(\ref{eq:deffjtP}) is a Lipschitz function whose best Lipschitz constant is $\| \bsy{\sigma}_{j}^{(\bsy{P})} \|_2$. 
\end{lem}

\begin{thm}\label{thm:concentratedProbIneq}
	Let $\bsy{g} \sim \mathcal{N}(0, \bsy{I}_{K \times K})$. For a fixed iteration $t$, let $\bsy{P} \in \mathcal{P}^{(t)}$ and $\bsy{R} = (\bsy{I} - \bsy{P}) \bsy{Y}$, \textit{i.e.}, $\bsy{R}$ is one of the residuals that could be generated by SOMP on the basis of $\bsy{Y}$ at iteration $t-1$ assuming that only correct atoms have been identified so far. Let
	\begin{equation}
	\Delta \mathbb{E}^{(t,\bsy{P})} := \mathbb{E}\left[ f_{j_c^{(t,\bsy{P})}}^{(t, \bsy{P})} (\bsy{g}) \right] - \max_{j \in \overline{\mathcal{S}}} \mathbb{E}\left[ f_{j}^{(t, \bsy{P})} (\bsy{g}) \right]
	\end{equation}
	and assume $\| \bsy{\phi}_j \|_2 = 1$ for $j \in \lbrack n \rbrack$. If $\Delta \mathbb{E}^{(t,\bsy{P})} > 0$, then the probability of SOMP making an incorrect decision when executing one iteration on $\bsy{R}$ is upper bounded by
	\begin{equation}
	(n - |\mathcal{S}| + 1) \exp \left[ - \dfrac{1}{8 \| \bsy{\sigma} \|_2^2} (\Delta \mathbb{E}^{(t,\bsy{P})})^2 \right].
	\end{equation}
\end{thm}

In Figure~\ref{fig:TheoremInterpretation}, it is observed that the mean of the folded normal distribution and the underlying normal distribution are virtually equal for sufficiently high values of $\sum_{k=1}^K \beta_{j,k}^{(t, \bsy{P})}$ while a discrepancy is observed for low values, as for the second incorrect atom in Figure \ref{fig:TheoremInterpretation}. As explained in Section~\ref{subsec:foldedNormVar}, taking the absolute value of a normal distribution yields a folding of the PDF and thereby increases the mean with regard to the underlying normal distribution. It implies that, for two atom indexes $j_1$ and $j_2$, $\gamma_{j_1}^{(t, \bsy{P})} > \gamma_{j_2}^{(t, \bsy{P})}$ is no guarantee for $\mathbb{E}\big[ f_{j_1}^{(t, \bsy{P})} (\bsy{g}) \big] > \mathbb{E}\big[ f_{j_2}^{(t, \bsy{P})} (\bsy{g}) \big]$ where $\gamma_{j}^{(t, \bsy{P})} := \sum_{k=1}^{K} \beta_{j,k}^{(t, \bsy{P})}$ denotes the value of SOMP metric for the $j$-th atom. We provide two explanations for this phenomenon:
\begin{itemize}
	\item The noise vectors $\bsy{\sigma}_{j}^{(\bsy{P})}$ may exhibit different $\ell_2$ norms for each atom. Thus, even for $K = 1$, a possibly higher noise level for the $j_2$-th atom and a sufficiently small gap $\gamma_{j_1}^{(t, \bsy{P})} - \gamma_{j_2}^{(t, \bsy{P})}$ might yield $\mathbb{E}\lbrack f_{j_1}^{(t, \bsy{P})} (\bsy{g}) \rbrack < \mathbb{E}\lbrack f_{j_2}^{(t, \bsy{P})} (\bsy{g}) \rbrack$ even though $\gamma_{j_1}^{(t, \bsy{P})} > \gamma_{j_2}^{(t, \bsy{P})}$. Note that this phenomenon cannot occur for iteration $0$ as the noise variances are necessarily equal in that case since $\bsy{P} = \bsy{0}$.
	\item For $K > 1$, and for possibly identical noise vectors $\bsy{\sigma}_{j}^{(\bsy{P})}$, the way the $\beta_{j,k}^{(t, \bsy{P})}$ are distributed for $k$ plays a significant role. For example, if only one of the $K$ quantities $\beta_{j_2,k}^{(t, \bsy{P})}$ is overwhelmingly greater than the others then very limited folding will occur on this entry while significant folding is present on the other ones. Conversely, if the entries of $\beta_{j_1,k}^{(t, \bsy{P})}$ are more uniformly distributed, then the overall increase in expectation due to the folding is decreased. Thus, depending on the way the $K$ terms $\beta_{j,k}^{(t, \bsy{P})}$ are distributed for $j_1$ and $j_2$, $\mathbb{E}\lbrack f_{j_1}^{(t, \bsy{P})} (\bsy{g}) \rbrack < \mathbb{E}\lbrack f_{j_2}^{(t, \bsy{P})} (\bsy{g}) \rbrack$ might hold despite having $\gamma_{j_1}^{(t, \bsy{P})} > \gamma_{j_2}^{(t, \bsy{P})}$.
\end{itemize}
Note that the reason why $\max_{j \in \overline{\mathcal{S}}} \mathbb{E}\big[ f_{j}^{(t, \bsy{P})} (\bsy{g}) \big]$ cannot be replaced by $\mathbb{E}\big[ f_{j_i^{(t, \bsy{P})}}^{(t, \bsy{P})} (\bsy{g}) \big]$ in the definition of $\Delta \mathbb{E}^{(t,\bsy{P})}$ is similar to the two explanations above.\\

As stated in the abstract, the two points above provide convincing explanations of why $\gamma_c^{(t,\bsy{P})} > \gamma_i^{(t,\bsy{P})}$, \textit{i.e.}, success in the noiseless case, is no guarantee for correct recovery when $K \rightarrow \infty$. Indeed, each random variable $(1/K) \mathbb{E}\big[ f_{j}^{(t, \bsy{P})} (\bsy{g}) \big]$ concentrates arbitrarily well around its expectation for a sufficiently high value of $K$. It is therefore to be expected that asymptotic correct recovery at iteration $t$ and for orthogonal projector $\bsy{P}$, \textit{i.e.}, recovery when $K \rightarrow \infty$, is ensured if and only if $\Delta \mathbb{E}^{(t,\bsy{P})} > 0$. The condition $\gamma_c^{(t,\bsy{P})} > \gamma_i^{(t,\bsy{P})}$ is actually neither necessary nor sufficient for correct asymptotic recovery as the phenomena discussed above might benefit the correct atoms, \textit{i.e.}, the condition is not necessary, or the incorrect atoms, \textit{i.e.}, the condition is not sufficient. However, the analyses that follow rely on the conservative assumption that only the incorrect atoms benefit from the increase of expectation due to the folding.
\section{Upper bound on the probability of SOMP failing during the first $s+1$ iterations}\label{sec:specificResults}

The final problem to be addressed is to derive an upper bound on the probability of SOMP picking incorrect atoms during the first $s+1$ iterations. In particular, the bound should not require to know the sequence of orthogonal projectors intervening at each iteration. To do so, we design a lower bound for $\Delta \mathbb{E}^{(t,\bsy{P})}$ in Theorem~\ref{thm:concentratedProbIneq} independent of the orthogonal projector $\bsy{P} \in \mathcal{P}^{(t)}$ and that conveys in a simple manner the impact on SOMP performance of
\begin{itemize}
	\item The sensing properties of the matrix $\bsy{\Phi}$.
	\item The absolute reliability of SOMP in the noiseless case, \textit{i.e.}, $\gamma_{c}^{(t, \bsy{P})} - \gamma_{i}^{(t, \bsy{P})}$, and the associated lower bound in Lemma~\ref{lem:noislessIneqDecisions}.
	\item The noise variances $\sigma_k^2$.
\end{itemize}
Once this bound is obtained, we will show that Theorem~\ref{thm:concentratedProbIneq} yields an upper bound on the probability of failure of SOMP from iteration $0$ to iteration $s < |\mathcal{\mathcal{S}}|$ included, \textit{i.e.}, an upper bound on the probability that SOMP picks at least one incorrect atom during the first $s+1$ iterations.
\subsection{Deriving the lower bound of $\Delta \mathbb{E}^{(t,\bsy{P})}$}\label{subsec:lowBoundDeltaE}
In this section, we propose a lower bound for $\mathbb{E}\lbrack f_{j_c^{(t,\bsy{P})}}^{(t, \bsy{P})} (\bsy{g}) \rbrack$ and an upper bound for $\max_{j \in \overline{\mathcal{S}}} \mathbb{E}\lbrack f_{j}^{(t, \bsy{P})} (\bsy{g}) \rbrack$, that are both valid for all $\bsy{P} \in \mathcal{P}^{(t)}$, so that a lower bound for $\Delta \mathbb{E}^{(t,\bsy{P})}$ is obtained. Exceptionally, the proof is presented in the core of the paper as doing so gives the reader a better understanding of the derivation method. Regarding $\mathbb{E}\lbrack f_{j_c^{(t,\bsy{P})}}^{(t, \bsy{P})} (\bsy{g}) \rbrack$, we have 
\begin{align*}
\mathbb{E}\left[ f_{j_c^{(t,\bsy{P})}}^{(t, \bsy{P})} (\bsy{g}) \right] & = \sum_{k=1}^K \mathbb{E}\left[ \left| \beta_{j_c^{(t,\bsy{P})},k}^{(t, \bsy{P})} + \sigma_{j_c^{(t,\bsy{P})}, k}^{(\bsy{P})} g_k \right|\right]\\
& \geq \sum_{k=1}^K \left| \beta_{j_c^{(t,\bsy{P})},k}^{(t, \bsy{P})} + \sigma_{j_c^{(t,\bsy{P})}, k}^{(\bsy{P})} \mathbb{E}\left[g_k\right] \right| \\
& = \sum_{k=1}^K \left| \beta_{j_c^{(t,\bsy{P})},k}^{(t, \bsy{P})} \right| =: \gamma_{c}^{(t, \bsy{P})}
\end{align*}
where the inequality results from Jensen's inequality for convex functions. Unless $\beta_{j_c^{(t,\bsy{P})},k}^{(t, \bsy{P})}/\sigma_{j_c^{(t,\bsy{P})}, k}^{(\bsy{P})} \rightarrow 0$, the bound above is not sharp as it discards the gain of expectation resulting from the folding. Moreover, the triangle inequality provides
\begin{align*}
\mathbb{E}\left[ f_{j}^{(t, \bsy{P})} (\bsy{g}) \right] & = \mathbb{E}\left[ \sum_{k=1}^K  \left| \beta_{j,k}^{(t, \bsy{P})} + \sigma_{j, k}^{(\bsy{P})} g_k \right|\right]\\
& \leq \sum_{k=1}^K \left| \beta_{j,k}^{(t, \bsy{P})} \right| + \sum_{k=1}^{K} \sigma_{j, k}^{(\bsy{P})} \mathbb{E}\left[ |g_k |\right]
\end{align*}
where $\mathbb{E}\lbrack |g_k |\rbrack = \sqrt{2/\pi}$ and $\sum_{k=1}^K | \beta_{j,k}^{(t, \bsy{P})} | \leq \sum_{k=1}^K | \beta_{j_i^{(t,\bsy{P})},k}^{(t, \bsy{P})} | = \gamma_{i}^{(t, \bsy{P})}$ for $j \in \overline{\mathcal{S}}$. The inequality is not sharp as it assumes maximum folding, \textit{i.e.}, the folding obtained if the underlying mean is $0$, for all the terms $| \beta_{j,k}^{(t, \bsy{P})} + \sigma_{j, k}^{(\bsy{P})} g_k |$ while, in the general case, at least one of the $\beta_{j,k}^{(t, \bsy{P})}$ should be non-zero. Also, the inequality $\sigma_{j, k}^{(\bsy{P})} \leq \sigma_k$ holds (see Section~\ref{subsec:distL1Norm}). As a conclusion, we obtain
\begin{equation}\label{eq:naiveIneqv1}
\Delta \mathbb{E}^{(t,\bsy{P})} \geq (\gamma_{c}^{(t, \bsy{P})} - \gamma_{i}^{(t, \bsy{P})}) - \sqrt{\dfrac{2}{\pi}} \| \bsy{\sigma} \|_1
\end{equation}
where an expression quantifying the absolute reliability of the decisions in the noiseless case intervenes, \textit{i.e.}, $\gamma_{c}^{(t, \bsy{P})} - \gamma_{i}^{(t, \bsy{P})}$, as well as a penalty depending on the noise standard deviations for all the $K$ channels, \textit{i.e.}, $\sqrt{2/\pi} \| \bsy{\sigma} \|_1$. Using Lemma~\ref{lem:noislessIneqDecisions} yields
\begin{equation*}
\Delta \mathbb{E}^{(t,\bsy{P})} \geq \left(1 - \dfrac{1}{\Gamma} \right) \psi  \tau_X - \sqrt{\dfrac{2}{\pi}} \| \bsy{\sigma} \|_1
\end{equation*}
where theoretical expressions for $\Gamma$, $\psi$, and $\tau_X$ are discussed in Section~\ref{sec:robustnessNoNoise}. Note that the result only holds whenever $\Gamma > 1$ for the theoretical expression that has been chosen. The next step is to combine the derived lower bound with Theorem~\ref{thm:concentratedProbIneq}.
\subsection{An upper bound on the probability of failure of SOMP during the first $s+1$ iterations}

The main difficulty here is to derive the desired upper bound not knowing the sequence of supports that are chosen during the first $s+1$ iterations. To circumvent this issue, all the possible orthogonal projection matrices $\bsy{P} \in \mathcal{P}^{(t)}$ are tested while only one intervenes in practice. This sub-optimal approach is linked to the fact that $\bsy{P}$ and $\bsy{E}$ are statistically dependent random variables as the noise contributes to determining which atom is picked and thus influences the orthogonal projection matrices. The statistical link between both variables appears to be difficult to capture and we consequently chose to use this workaround. This ``trick'' has been used in similar theoretical analyses \cite{gribonval2008atoms, eldar2010average} and, to the best of the authors' knowledge, a better solution has yet to be found in the literature. The proof is available in the Appendix.

\begin{thm}\label{thm:finalThmNew}
	Using the quantities $\Gamma$, $\psi$, and $\tau_X$ defined in Equation~(\ref{eq:DefGamma}) and (\ref{eq:DefTauX}), let us consider
	\begin{equation}
	\Delta \mathbb{E} := \left(1 - \dfrac{1}{\Gamma} \right) \psi \tau_X - \sqrt{\dfrac{2}{\pi}} \| \bsy{\sigma} \|_1
	\end{equation}
	and assume $\| \bsy{\phi}_j \|_2 = 1$ for $j \in \lbrack n \rbrack$. If $\Delta \mathbb{E} > 0$, then the probability of SOMP picking at least one incorrect atom from iteration $0$ to iteration $s$ included is upper bounded by
	\begin{equation}
	n \mathcal{C}_s \exp \left[ - \dfrac{1}{8 \| \bsy{\sigma} \|_2^2} (\Delta \mathbb{E})^2 \right]
	\end{equation}
	where $\mathcal{C}_s := \sum_{t=0}^{s} \binom{|\mathcal{S}|}{t} \stackrel{(s \geq 1)}{\leq} \left( \frac{e (|\mathcal{S}|+s-1)}{s} \right)^s + 1$.
\end{thm}
The expressions appearing in Theorem~\ref{thm:finalThmNew} can be deciphered in the following manner:
\begin{itemize}
	\item  As explained in Section~\ref{sec:robustnessNoNoise}, the term $(1 - 1/\Gamma) \psi \tau_X$ conveys the impact of the reliability of the decisions prior to the addition of noise.
	\item Both terms $\sqrt{2/\pi} \| \bsy{\sigma} \|_1$ and $1/\| \bsy{\sigma} \|_2^2$ convey the negative influence of the noise on SOMP performance. The phenomena they account for are however different in nature. The quantity $1/\| \bsy{\sigma} \|_2^2$ translates the spread of the PDF of each $f_{j}^{(t, \bsy{P})}$ which is characterized by the spread of the Gaussian-like functions in Figure~\ref{fig:TheoremInterpretation}. On the other hand, the expression $\sqrt{2/\pi} \| \bsy{\sigma} \|_1$ describes the negative impact that the noise has on the existing reliability in the noiseless case, \textit{i.e.}, $\gamma_{c}^{(t, \bsy{P})} - \gamma_{i}^{(t, \bsy{P})}$ in Equation~(\ref{eq:naiveIneqv1}). Our commentary of Theorem~\ref{thm:concentratedProbIneq} thoroughly discusses these matters.
\end{itemize}
The next section deals with the understanding of SOMP performance whenever $K$ increases. In particular, the AERC mentioned in the introduction will be derived.%
\subsection{Probability of failure for increasing values of $K$}

To gain further insight into what Theorem~\ref{thm:finalThmNew} implies when $K$ increases, we are using the quantities defined in the introduction, \textit{i.e.}, $\mu_X(K) :=  \min_{j \in \mathcal{S}} \frac{1}{K} \sum_{k=1}^{K} | X_{j,k} |$, $\sigma(K)^2 = \frac{1}{K} \sum_{k=1}^K \sigma_k^2$, $\omega_{\sigma} = \max_{1 \leq K < \infty} (1/\sqrt{K}) \| \bsy{\sigma} \|_1/\| \bsy{\sigma} \|_2$ and $\mathrm{SNR}_{\mathrm{min}} := \min_{1 \leq K < \infty}  \frac{\mu_X(K)}{\sigma(K)}$. The definition of $\sigma(K)^2$  implies that $\| \bsy{\sigma} \|_2^2 = K \sigma(K)^2$. Thus, using the expression of $\tau_X$ provided by Lemma~\ref{lem:lowBNoiselessMetricCorrectAtoms}, we obtain
\begin{align*}
\dfrac{\Delta \mathbb{E}}{\| \bsy{\sigma} \|_2} & = \sqrt{K} \left( \left(1 - \frac{1}{\Gamma} \right) \psi  \frac{\mu_X(K)}{\sigma (K)} - \sqrt{\dfrac{2}{K \pi}} \dfrac{\| \bsy{\sigma} \|_1}{\| \bsy{\sigma} \|_2} \right)\\
& \geq \sqrt{K} \left( \left(1 - \frac{1}{\Gamma} \right) \psi  \mathrm{SNR}_{\mathrm{min}} - \sqrt{\dfrac{2}{\pi}} \omega_{\sigma} \right)
\end{align*}
Hence, ensuring $\xi > 0$ yields $(\Delta \mathbb{E})^2 / \| \bsy{\sigma} \|_2^2 \geq K \xi^2$ where
\begin{equation}\label{eq:xiDef}
\xi := \left(1 - \dfrac{1}{\Gamma} \right) \psi  \mathrm{SNR}_{\mathrm{min}} - \sqrt{\dfrac{2}{\pi}} \omega_{\sigma}.
\end{equation}
As a consequence, under the condition $\xi > 0$, the probability that SOMP picks at least one incorrect atom during the first $s+1$ iterations is upper bounded by 
\begin{equation}
n \mathcal{C}_s \exp \left[ - \dfrac{1}{8} K \xi^2 \right]
\end{equation}
which indicates that the condition $\xi > 0$ is sufficient for asymptotic recovery, \textit{i.e.}, it is a valid AERC. More precisely, the expression of $\xi$ shows that  $\mathrm{SNR}_{\mathrm{min}}$ should be sufficiently high to guarantee the asymptotic recovery. It is also worth noticing that, for $\mathrm{SNR}_{\mathrm{min}} \rightarrow \infty$, our result is equivalent to the exact recovery criterion (ERC) $1 - 1/\Gamma > 0$ in the noiseless case.\\

To conclude the technical discussions, we show that, as long as $\xi > 0$, there always exists a value of $K$ ensuring an arbitrary maximum probability of failure for SOMP. Fixing the maximum probability of failure $p_{\mathrm{err}} > 0$, elementary algebraic operations show that satisfying $K > K_{\mathrm{min}}(p_{\mathrm{err}})$ yields a probability of failure inferior to $p_{\mathrm{err}}$ where
\begin{equation}\label{eq:Kmin}
K_{\mathrm{min}}(p_{\mathrm{err}}) := \dfrac{8}{\xi^2} \log \left(\dfrac{n \mathcal{C}_s}{p_{\mathrm{err}}} \right).
\end{equation}
The quantity $\xi$ increases with $\mathrm{SNR}_{\mathrm{min}}$ so that the number of sparse signals $K$ needed to achieve a prescribed maximum probability of failure decreases whenever the SNR improves. \\ 

Finally, according to Theorem~\ref{thm:finalThmNew} and for $s \geq 1$, we have $\mathcal{C}_s \leq \left( \frac{e (|\mathcal{S}|+s-1)}{s} \right)^s + 1 \lesssim \left( \frac{e (|\mathcal{S}|+s-1)}{s} \right)^s$ so that $\log \mathcal{C}_s \lesssim s  \log e \frac{|\mathcal{S}|+s-1}{s} < s \log \frac{ 2 e |\mathcal{S}|}{s}$. Equation~(\ref{eq:Kmin}) now yields
\begin{equation*}
K_{\mathrm{min}}(p_{\mathrm{err}}) \lesssim \dfrac{1}{\xi^2} \left( \log n + s \log \frac{ 2 e |\mathcal{S}|}{s} - \log p_{\mathrm{err}} \right),
\end{equation*}
which suggests that $K_{\mathrm{min}}$ should scale logarithmically with $1/p_{\mathrm{err}}$ and the number of atoms $n$ and linearly with the number of iterations to be performed $s+1$.

\section{Numerical results} \label{sec:numRes}
\subsection{Objective and methodology}\label{subsec:numResObjMethod}
Through numerical simulations, we wish to validate that our theoretical developments properly describe SOMP performance. To do so, we show that performing some adjustments in Equation~(\ref{eq:Kmin}) enables us to accurately predict the minimum number of sparse signals $K$ needed to achieve a prescribed probability of error $p_{\mathrm{err}}$ when using SOMP to perform the full support recovery. More precisely, we consider three parameters $\alpha$, $\beta$, and $\gamma$ to be identified so that Equation~(\ref{eq:Kmin}) rewrites 
\begin{equation}\label{eq:KminvNumRes}
K_{\mathrm{min}}(p_{\mathrm{err}}) := \dfrac{8}{\left(\alpha \; \mathrm{SNR}_{\mathrm{min}} - \omega_{\sigma} \beta\right)^2} \left( \gamma - \log p_{\mathrm{err}} \right).
\end{equation}
The parameter $\alpha := (1- 1/\Gamma)\psi$ is related to the relative reliability of SOMP in the noiseless case. In theory, the quantity $\beta$  is equal to $\sqrt{2/\pi}$. However, the theoretical development provides values that are not sharp in the general case (see our discussion in Section~\ref{subsec:lowBoundDeltaE}) and possibly even less for more constrained signal models. Finally, the parameter $\gamma = \log(n \mathcal{C}_{|\mathcal{S}|-1})$ conveys the impact of the number of atoms $n$ and the number of iterations, \textit{i.e.}, $|\mathcal{S}|$ in this case. In practice, only a few of the $n -|\mathcal{S}|$ incorrect atoms have a non-negligible probability to be picked so that $n$ should be replaced by $\overline{n} \ll n - |\mathcal{S}|$. Also, as we discussed before, $\mathcal{C}_{|\mathcal{S}|-1}$ is a suboptimal term that results from the fact that we considered all the possible correct supports at each iteration to deduce the probability of error while, in practice, only one support out of the numerous possibilities matters. The identification procedure uses the cost function 
\begin{equation*}
\sum_{\substack{p_{\mathrm{err}} \in \mathcal{I}_{\mathrm{perr}}\\ {\mathrm{SNR}_{\mathrm{min}} \in \mathcal{I}_{\mathrm{SNR}}}}} \left[ \sqrt{K_{\mathrm{min}}} (\alpha \mathrm{SNR}_{\mathrm{min}} - \beta \omega_{\sigma})  - \sqrt{8 (\gamma - \log p_{\mathrm{err}})} \right]^2
\end{equation*}
where $K_{\mathrm{min}}$ is obtained by means of simulations, $\mathcal{I}_{\mathrm{perr}} := \lbrace 0.05; 0.5; 0.9 \rbrace$ and $\mathcal{I}_{\mathrm{SNR}}$ is to be specified afterwards. The cost function is evaluated for each $3$-tuple $(\alpha, \beta, \gamma) \in \mathcal{I}_{\alpha} \times \mathcal{I}_{\beta} \times \mathcal{I}_{\gamma}$ where each set $\mathcal{I}_{\alpha}$, $\mathcal{I}_{\beta}$ and $\mathcal{I}_{\gamma}$ consists of $500$ uniformly distributed points in the intervals $\lbrack 0.1; 1.4 \rbrack$, $\lbrack 0; \sqrt{2/\pi} \rbrack$ and $\lbrack 0; 5 \rbrack$, respectively. The $3$-tuple $(\alpha, \beta, \gamma)$ minimizing the cost function is then chosen.

\subsection{Simulations signal model}

We now describe the signal model we consider to run the simulations. It is more simple than the general model described in Section~\ref{subsec:sigModObj} so that $\mathrm{SNR}_{\mathrm{min}}$ can be easily computed. First of all, the noise standard deviation vector $\bsy{\sigma}$ has its odd-indexed (even-indexed) entries identical, \textit{i.e.}, $\bsy{\sigma} = (\sigma_{\mathrm{odd}}, \sigma_{\mathrm{even}}, \sigma_{\mathrm{odd}}, \sigma_{\mathrm{even}}, \dots)$ where we define $r_{\sigma} = \sigma_{\mathrm{even}} / \sigma_{\mathrm{odd}}$. It is then easy to show that $\omega_{\sigma} = \frac{1}{\sqrt{2}} (r_{\sigma}+1)/ \sqrt{r_{\sigma}^2+1}$ for $K$ even and/or $r_{\sigma} = 1$.  Furthermore, we assume $X_{j,k} = \varepsilon_{j,k} \mu_X$ ($j \in \mathcal{S}, k \in \lbrack K \rbrack$) where $\mu_X$ is fixed and the $\lbrace \varepsilon_{j,k} \rbrace_{j \in \lbrack n \rbrack, k \in \lbrack K \rbrack}$ are statistically independent Rademacher variables, \textit{i.e.}, random variables returning either $-1$ or $1$ with probability $0.5$ for both outcomes. Under this assumption, we have $\mu_X(K) = \mu_X$ for every $K$.
\subsection{Simulation setup}
The sensing matrix $\bsy{\Phi}$ has been chosen of size $250 \times 1000$, \textit{i.e.}, $m = 250$ and $n = 1000$, and its columns are the realizations of statistically independent vectors uniformly distributed on the unit hypersphere $\mathbb{S}^{m-1} := \lbrace \bsy{\phi} \in \mathbb{R}^m : \| \bsy{\phi} \|_2 = 1 \rbrace$. The sensing matrix is identical for all the simulations that have been conducted. All the simulations will be such that $\sigma (K) = 1$. The reason is that it is always possible to recast a signal model for which $\sigma (K) = \zeta \neq 1$ as another one satisfying $\sigma (K) = 1$ while maintaining the value of $\mathrm{SNR}_{\mathrm{min}}$ by multiplying $\bsy{Y}$ by $1/\zeta$. This multiplication does not affect SOMP decisions as all the inner products intervening in SOMP decision metric are equally affected (see step $4$ in Algorithm~\ref{alg:SOMP}). Note that our results and our Matlab software are available in \cite{determe2016SOMPSoftwarePackage}.

\subsection{Results and analysis}

\begin{figure}[H]
	\centering
	\subfloat[Configuration with $|\mathcal{S}| = 10$ -- Identified parameters (with $\mathcal{I}_{\mathrm{SNR}} = \lbrace 1.25; 1.5; 1.75 \rbrace$, see Section~\ref{subsec:numResObjMethod}): $\alpha = 1.0535$, $\beta = 0.54045$, and $\gamma = 2.0741$. \label{subfig:numResCardS10}]{%
		\includegraphics[scale=0.75]{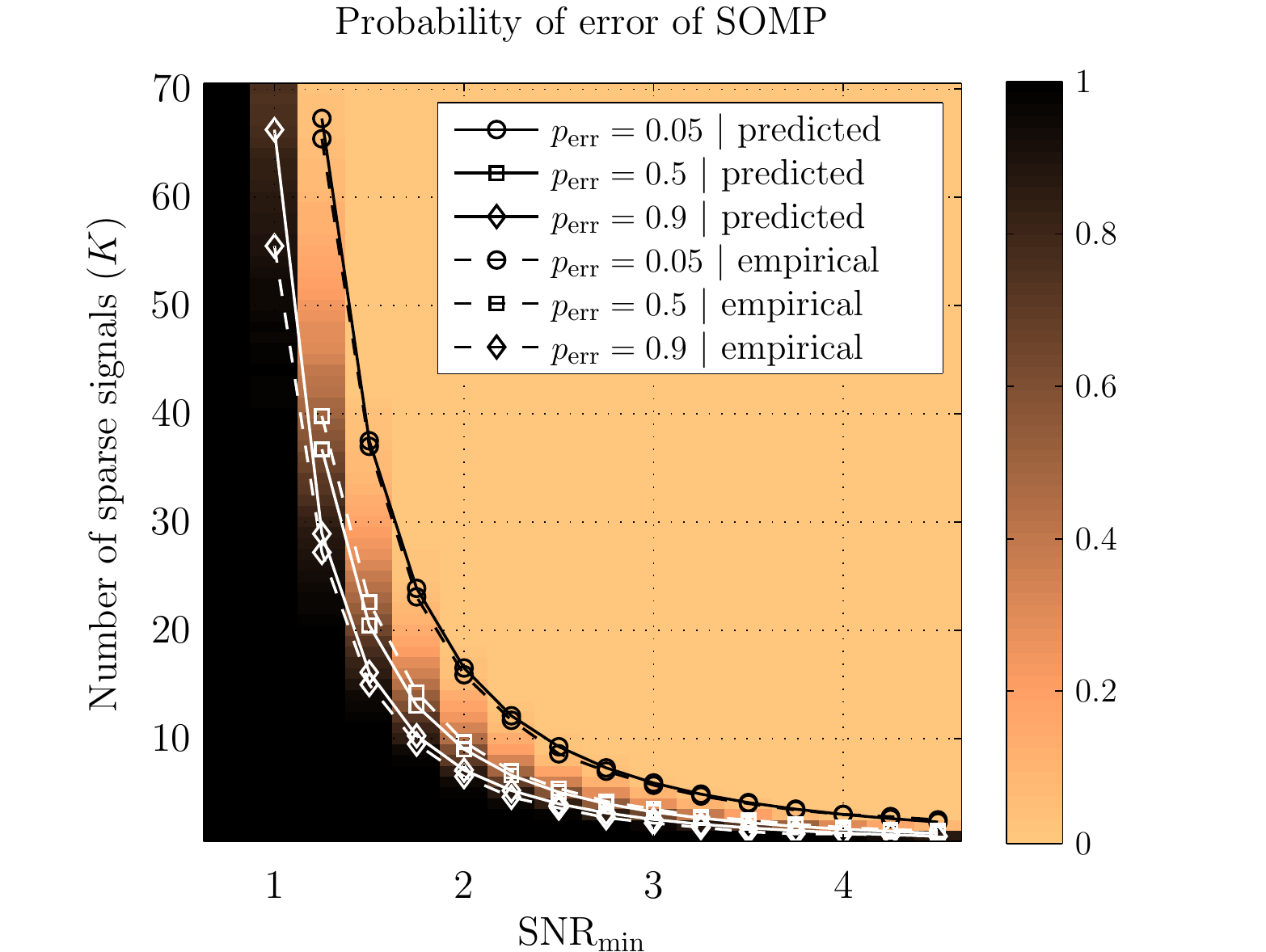}
	}
	
	\subfloat[Configuration with $|\mathcal{S}| = 20$ -- Identified parameters (with $\mathcal{I}_{\mathrm{SNR}} = \lbrace 1.5; 1.75; 2 \rbrace$, see Section~\ref{subsec:numResObjMethod}): $\alpha = 1.0535$, $\beta = 0.58682$, and $\gamma = 2.5451$. \label{subfig:numResCardS20}]{%
		\includegraphics[scale=0.75]{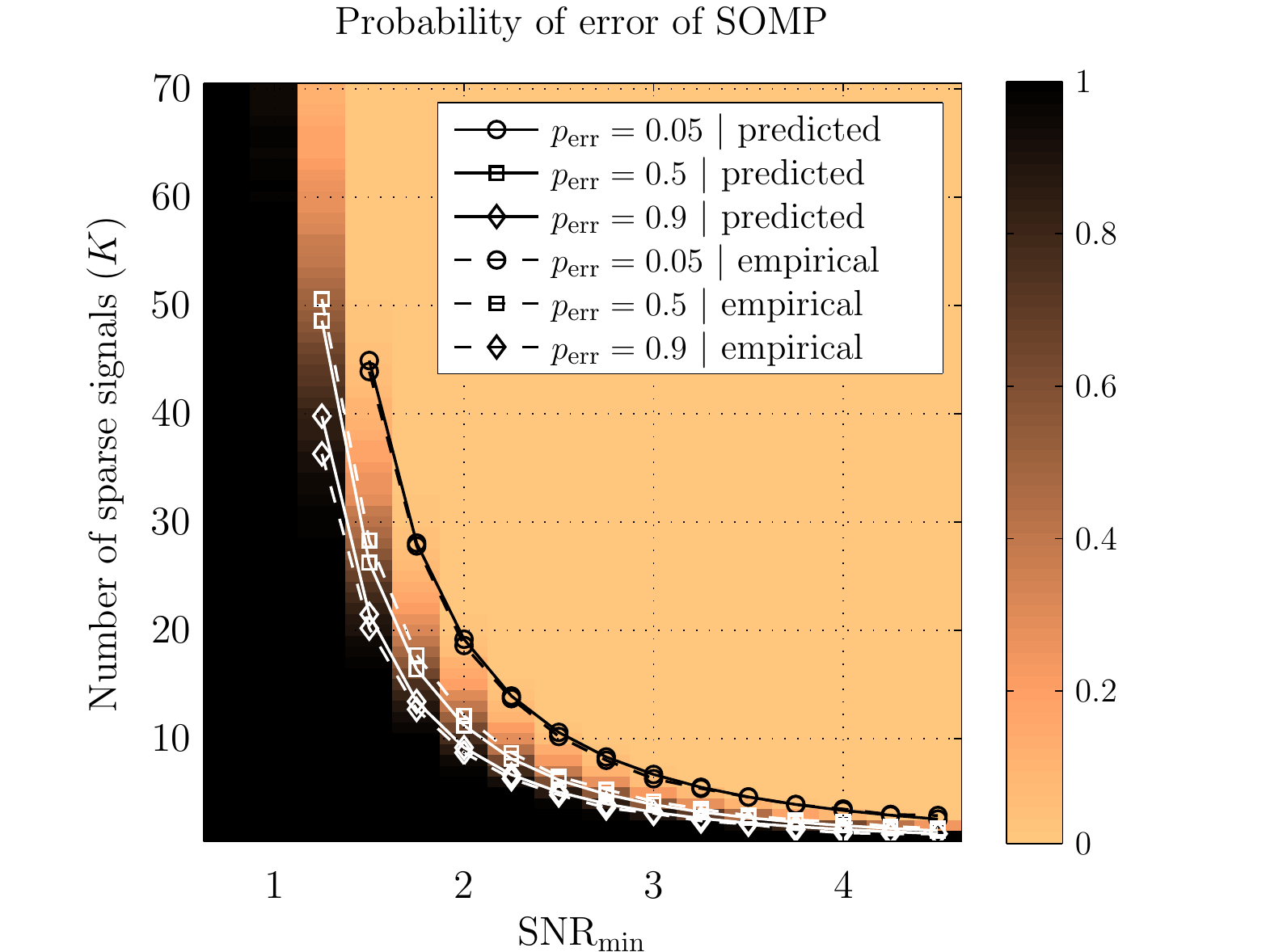}
	}
	\caption{Probability of SOMP comitting at least one error when performing the joint full support recovery -- The continuous curves plot the values of $K_{\mathrm{min}}$ predicted by Equation~(\ref{eq:KminvNumRes}) for several values of $p_{\mathrm{err}}$ when using the identified parameters $\alpha$, $\beta$, and $\gamma$. The dashed curves correspond to the numerical level sets for a fixed probability of failure $p_{\mathrm{err}}$ corresponding to that of the associated continuous curve. Note that the color of each curve has no particular meaning as its only purpose is constrast enhancement. The number of Monte-carlo cases is equal to $2000$ for each $2$-tuple $(\mathrm{SNR}_{\mathrm{min}}, K)$.}
	\label{fig:numResHomosced}
\end{figure}

We wish to determine whether Equation~(\ref{eq:KminvNumRes}) accurately predicts SOMP performance provided that the parameters $\alpha$, $\beta$, and $\gamma$ are properly identified. To do so, we begin with Monte-Carlo simulations for the homoscedastic signal model, \textit{i.e.}, for $r_{\sigma} = 1$, for two different support cardinalities, \textit{i.e.}, $|\mathcal{S}| = 10$ and $|\mathcal{S}| = 20$. For each support cardinality, we identify a $3$-tuple of parameters $(\alpha, \beta, \gamma)$ as explained in Section~\ref{subsec:numResObjMethod}. The predictions based on the identified parameters are then compared against another set of Monte-Carlo simulations for which $r_{\sigma}$ varies. For each Monte-Carlo simulation, the support is chosen at random and the random variables $\bsy{X}$ and $\bsy{E}$ are statistically independent. Figure~\ref{fig:numResHomosced} plots the results obtained for the homoscedastic signal model.\\

As observed in Figure~\ref{fig:numResHomosced}, the fitting of Equation~(\ref{eq:KminvNumRes}) to the numerical results is satisfactory over the full range of the simulations. The identification procedure has yielded values for $\alpha$ slightly higher than $1$, which is incoherent with the theory as the value $1$ is obtained whenever $\Gamma \rightarrow \infty$ and $\delta_{|\mathcal{S}|} = 0$. Also, the values of $\beta$ are similar for both cardinalities of $\mathcal{S}$ and are lower than $\sqrt{2/\pi} \simeq 0.7979$, as predicted by the theory. The values of $\alpha$ and $\beta$ remain high but the constant $8$ in Equation~(\ref{eq:KminvNumRes}) might not be sharp. In particular, if $8$ is replaced by $8/\theta$, then Equation~(\ref{eq:KminvNumRes}) provides identical values of $K_{\mathrm{min}}$ provided that $\alpha$ becomes $\alpha/\sqrt{\theta}$ and $\beta$ is replaced by $\beta/\sqrt{\theta}$. The values obtained for $\gamma$ are equal to $2.0741$ and $2.5451$, which indicates that $n \mathcal{C}_{|\mathcal{S}|-1}$ should be replaced by $\exp(2.0741) \simeq 7.9574$ and $\exp(2.5451) \simeq 12.7445$, respectively. The higher value of $\gamma$ for $|\mathcal{S}| = 20$ is coherent with the theory as $\mathcal{C}_{|\mathcal{S}|-1}$ increases with $|\mathcal{S}|$. The values obtained for $n \mathcal{C}_{|\mathcal{S}|-1}$ also suggest that the expression $n \mathcal{C}_{|\mathcal{S}|-1}$ is not sharp.\\

\begin{figure}[!ht]
	\centering
	\includegraphics[scale=0.85]{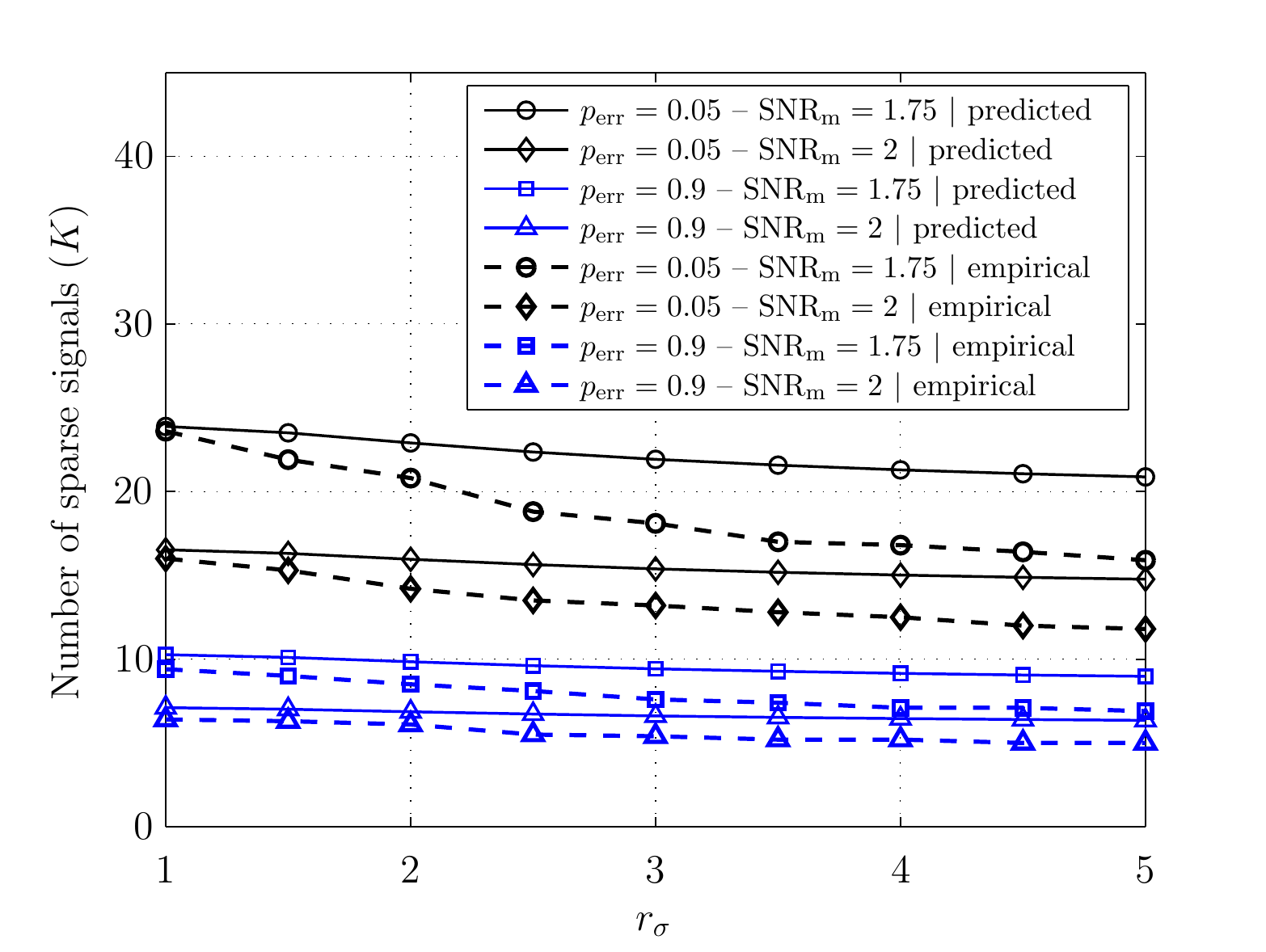}
	\caption{Levels sets of the probability of SOMP comitting at least one error when performing the joint full support recovery -- $|\mathcal{S}| = 10$ -- The continuous curves plot the values of $K_{\mathrm{min}}$ predicted by Equation~(\ref{eq:KminvNumRes}) for several $2$-tuples $(p_{\mathrm{err}}, \mathrm{SNR}_{\mathrm{min}})$ when using the  parameters identified in Figure~\ref{subfig:numResCardS10}. The dashed curves correspond to the numerical level sets corresponding to that of the associated continuous curve. The number of Monte-carlo cases is equal to $2000$ for each $2$-tuple $(r_{\sigma}, K)$. Only even values of $K$ have been simulated.}
	\label{fig:numResRatioSigmaCardS10}
\end{figure}

Figure~\ref{fig:numResRatioSigmaCardS10} plots the results obtained for $|\mathcal{S}| = 10$ when $r_{\sigma}$ increases. While our theoretical developments appropriately predict that the value of $K_{\mathrm{min}}$ decreases as the noise standard deviation vector $\bsy{\sigma}$ gets sparser, \textit{i.e.}, as $r_{\sigma}$ increases, our model is pessimistic with regard to the amplitude of the improvement of $K_{\mathrm{min}}$. We hypothesize that the inequalities used in Section~\ref{subsec:lowBoundDeltaE} are not sharp enough to accurately predict $K_{\mathrm{min}}$ on the basis of $r_{\sigma}$. The identified parameters $\alpha$, $\beta$, and $\gamma$ might also not be optimal.

\section{Conclusion}

In this paper, a theoretical analysis of SOMP operating in the presence of Gaussian additive noise has been presented. It has been shown that the signal to be recovered should be sufficiently large when compared to the mean noise power over all the measurement channels to succeed in the support recovery. Assuming this condition is met, the minimum number of sparse signals $K$ to be gathered to achieve a prescribed probability of failure has been derived. An interesting corollary of the aforementioned results is that ensuring SNR values above a threshold allows asymptotic recovery, \textit{i.e.}, the probability of error tends to $0$ as $K$ tends to infinity. Finally, numerical results confirmed the validity of the theoretical developments.

\appendix[Technical proofs]\label{appendix:proofs}

\subsection{Lemma~\ref{lem:lowBNoiselessMetricCorrectAtoms} example (Section~\ref{subsec:LBGammac})} 	\label{subsec:LBNoiselessMetricCorrectAtomsExample}

Let $|\mathcal{S}| \geq 2$, $\eta \in \mathcal{S}$, $K=1$, and $t=0$. Let us consider $\chi := \max_{j \in \mathcal{S} \setminus \lbrace \eta \rbrace}| \langle \bsy{\phi}_j, \bsy{\Phi}_{\mathcal{S}} \bsy{x}_{\mathcal{S}} \rangle | = \| \bsy{\Phi}_{\mathcal{S} \setminus \lbrace \eta \rbrace}^{\mathrm{T}} \bsy{\Phi}_{\mathcal{S}} \bsy{x}_{\mathcal{S}} \|_{\infty}$, which is the maximal value of SOMP metric for all the correct atoms except $\bsy{\phi}_{\eta}$. As $\bsy{\Phi}_{\mathcal{S}} \bsy{x}_{\mathcal{S}} =\bsy{\Phi}_{\mathcal{S} \setminus \lbrace \eta \rbrace} \bsy{x}_{\mathcal{S} \setminus \lbrace \eta \rbrace} + \bsy{\phi}_{\eta} x_{\eta}$, we easily show that 
\begin{equation*}
\bsy{\Phi}_{\mathcal{S} \setminus \lbrace \eta \rbrace}^{\mathrm{T}} \bsy{\Phi}_{\mathcal{S} \setminus \lbrace \eta \rbrace} \bsy{x}_{\mathcal{S} \setminus \lbrace \eta \rbrace} = - \bsy{\Phi}_{\mathcal{S} \setminus \lbrace \eta \rbrace}^{\mathrm{T}} \bsy{\phi}_{\eta} x_{\eta}
\end{equation*}
implies $\chi=0$. Thus, enforcing $\chi=0$ is possible by setting $\bsy{x}_{\mathcal{S} \setminus \lbrace \eta \rbrace} = - (\bsy{\Phi}_{\mathcal{S} \setminus \lbrace \eta \rbrace}^{\mathrm{T}} \bsy{\Phi}_{\mathcal{S} \setminus \lbrace \eta \rbrace})^{-1} \bsy{\Phi}_{\mathcal{S} \setminus \lbrace \eta \rbrace}^{\mathrm{T}} \bsy{\phi}_{\eta} x_{\eta} = - \bsy{\Phi}_{\mathcal{S} \setminus \lbrace \eta \rbrace}^{+} \bsy{\phi}_{\eta} x_{\eta}$, which is a non-zero vector except for pathological cases. As a result, at iteration $t=0$, the noiseless SOMP metric might be non-zero  for a single correct atom only, \textit{i.e.}, $\bsy{\phi}_{\eta}$ in this particular example. 

\subsection{Proof of Theorem~\ref{thm:basicProbIneq}}

To simplify and shorten notations, we will abbreviate $f_j^{(t,\bsy{P})} = f_j^{(t,\bsy{P})} (\bsy{g})$ in this proof. We will consider that only the correct atom whose index is $j_c^{(t,\bsy{P})}$ has a chance to be picked. All the other correct atoms could be picked but the conservative analysis we present does not take into account this possibility. A sufficient condition to pick an atom belonging to $\mathcal{S}$ at iteration $t$ given the projection matrix $\bsy{P}$ is thus given by $\max_{j \in \overline{\mathcal{S}}} f_{j}^{(t,\bsy{P})} < f_{j_c^{(t,\bsy{P})}}^{(t,\bsy{P})}$ which is different from the condition $\max_{j \in \overline{\mathcal{S}}} f_{j}^{(t,\bsy{P})} < \max_{j \in S} f_{j}^{(t,\bsy{P})}$ where all the correct atoms could be picked. Denoting by $E_{\mathrm{succ}}^{(t, \bsy{P})}$ the event occurring when SOMP picks a correct atom at iteration $t$ given the projection matrix $\bsy{P} \in \mathcal{P}^{(t)}$,  we have 
\begin{equation*}
\mathbb{P} \lbrack E_{\mathrm{succ}}^{(t, \bsy{P})}  \rbrack \geq \mathbb{P} \left\lbrack f_{j_c^{(t, \bsy{P})}}^{(t,\bsy{P})} > \max_{j \in \overline{\mathcal{S}}} f_{j}^{(t, \bsy{P})}  \right\rbrack.
\end{equation*}
Let $\alpha > 0$, then the event $f_{j_c^{(t, \bsy{P})}}^{(t,\bsy{P})} > \max_{j \in \overline{\mathcal{S}}} f_{j}^{(t, \bsy{P})}$ is implied by $(\max_{j \in \overline{\mathcal{S}}} f_{j}^{(t, \bsy{P})} < \alpha) \cap (f_{j_c^{(t, \bsy{P})}}^{(t,\bsy{P})} > \alpha)$ so that a new lower bound is given by
\begin{align*}
\mathbb{P} \lbrack E_{\mathrm{succ}}^{(t, \bsy{P})}  \rbrack   & \geq \mathbb{P} \lbrack (\max_{j \in \overline{\mathcal{S}}} f_{j}^{(t, \bsy{P})} < \alpha) \cap (f_{j_c^{(t, \bsy{P})}}^{(t,\bsy{P})} > \alpha) \rbrack \\
& = 1 - \mathbb{P} \lbrack (\max_{j \in \overline{\mathcal{S}}} f_{j}^{(t, \bsy{P})} \geq \alpha) \cup (f_{j_c^{(t, \bsy{P})}}^{(t,\bsy{P})} \leq \alpha) \rbrack
\end{align*}
where the second line is obtained by considering the complementary event of that of the previous line. Using the union bound yields 
\begin{equation*}
\mathbb{P} \lbrack E_{\mathrm{succ}}^{(t, \bsy{P})}  \rbrack \geq 1 - \left( \mathbb{P} \left\lbrack f_{j_c^{(t, \bsy{P})}}^{(t,\bsy{P})} \leq \alpha  \right\rbrack + \mathbb{P} \left\lbrack \max_{j \in \overline{\mathcal{S}}} f_{j}^{(t, \bsy{P})} \geq \alpha  \right\rbrack \right).
\end{equation*}
Since the event $\max_{j \in \overline{\mathcal{S}}} f_{j}^{(t, \bsy{P})} \geq \alpha$ is equal to $\cup_{j \in \overline{\mathcal{S}}} \big[ f_{j}^{(t, \bsy{P})} \geq \alpha \big]$, using the union bound a second time yields 
\begin{equation*}
\mathbb{P} \big[ \max_{j \in \overline{\mathcal{S}}} f_{j}^{(t, \bsy{P})} \geq \alpha \big] \leq \sum_{j \in \overline{\mathcal{S}}} \mathbb{P} \big[ f_{j}^{(t, \bsy{P})} \geq \alpha \big].
\end{equation*}
Noticing that the probability of SOMP failing at iteration $t$ is equal to $1 - \mathbb{P} \lbrack E_{\mathrm{succ}}^{(t, \bsy{P})} \rbrack$ concludes the proof.

\subsection{Proof of Lemma~\ref{lem:fisLipschitz}}

We consider two arbitrary vectors $\bsy{x}, \bsy{y} \in \mathbb{R}^K$. Using sequentially the triangle inequality, the reverse triangle inequality, and the Cauchy–Schwarz inequality yields
\begin{equation*}
\begin{aligned}
& \left| f_{j}^{(t, \bsy{P})}(\bsy{x}) - f_{j}^{(t, \bsy{P})}(\bsy{y}) \right| \\
& = \left|  \sum_{k=1}^{K} \left( \left| \beta_{j,k}^{(t, \bsy{P})} + \sigma_{j, k}^{(\bsy{P})} x_k \right| - \left| \beta_{j,k}^{(t, \bsy{P})} + \sigma_{j, k}^{(\bsy{P})} y_k \right| \right) \right| \\
& \leq  \sum_{k=1}^{K} \left| \left| \beta_{j,k}^{(t, \bsy{P})} + \sigma_{j, k}^{(\bsy{P})} x_k \right| - \left| \beta_{j,k}^{(t, \bsy{P})} + \sigma_{j, k}^{(\bsy{P})} y_k \right| \right| \\
& \leq \sum_{k=1}^{K} \sigma_{j, k}^{(\bsy{P})} \left| x_k - y_k \right| = \left\langle \bsy{\sigma}_{j}^{(\bsy{P})}, (| x_k - y_k |)_{k \in [ K ]} \right\rangle \\
& \leq \| \bsy{\sigma}_{j}^{(\bsy{P})} \|_2 \| (| x_k - y_k |)_{k \in [ K ]} \|_2 = \| \bsy{\sigma}_{j}^{(\bsy{P})} \|_2 \| \bsy{x} - \bsy{y} \|_2.
\end{aligned}
\end{equation*}
The quantity $\| \bsy{\sigma}_{j}^{(\bsy{P})} \|_2$ is thus a valid Lipschitz constant and it is also the best one since setting $\bsy{x} = 1.5\; \bsy{\sigma}_{j}^{(\bsy{P})}$ and $\bsy{y} = 0.5\; \bsy{\sigma}_{j}^{(\bsy{P})}$ saturates the inequality as $\beta_{j,k}^{(t, \bsy{P})} \geq 0$ for all $k \in \lbrack K \rbrack$.

\subsection{Proof of Theorem~\ref{thm:concentratedProbIneq}}
For better readability we will omit the dependence on $\bsy{g}$ and abbreviate $f_j^{(t,\bsy{P})} = f_j^{(t,\bsy{P})} (\bsy{g})$. Since $f_{j}^{(t, \bsy{P})}$ is Lipschitz with a Lipschitz constant $L$ equal to $\| \bsy{\sigma}_{j}^{(\bsy{P})} \|_2$, the concentration inequalities in Section~\ref{subsec:LipschtizDef} yield, for $\varepsilon > 0$,
\begin{equation*}
\mathbb{P} \left(f_{j}^{(t, \bsy{P})} \geq  \mathbb{E}\lbrack f_{j}^{(t, \bsy{P})} \rbrack + \varepsilon \right) \leq \exp \left( \dfrac{-\varepsilon^2}{2 L^2}\right),
\end{equation*}
which rewrites (with $\alpha := \mathbb{E}\lbrack f_{j}^{(t, \bsy{P})} \rbrack + \varepsilon$)
\begin{equation*}
\mathbb{P} \left(f_{j}^{(t, \bsy{P})} \geq  \alpha \right) \leq \exp \left( - \dfrac{1}{2 L^2} \left(\alpha - \mathbb{E}\lbrack f_{j}^{(t, \bsy{P})} \rbrack  \right)^2\right)
\end{equation*}
provided that $\alpha > \mathbb{E}\lbrack f_{j}^{(t, \bsy{P})} \rbrack$. Similarly, using the concentration inequality obtained for $- f_{j}^{(t, \bsy{P})}$ yields
\begin{equation*}
\mathbb{P} \left(f_{j}^{(t, \bsy{P})} \leq  \alpha \right) \leq \exp \left( - \dfrac{1}{2 L^2} \left(\mathbb{E}\lbrack f_{j}^{(t, \bsy{P})} \rbrack - \alpha  \right)^2\right)
\end{equation*}
where $\alpha := \mathbb{E}\lbrack f_{j}^{(t, \bsy{P})} \rbrack - \varepsilon < \mathbb{E}\lbrack f_{j}^{(t, \bsy{P})} \rbrack$. \\

As suggested by Figure~\ref{fig:TheoremInterpretation}, the value of $\alpha$ is chosen in between $\max_{j \in \overline{\mathcal{S}}} \mathbb{E}\big[ f_{j}^{(t, \bsy{P})}  \big]$ and $\mathbb{E}\big[ f_{j_c^{(t,\bsy{P})}}^{(t, \bsy{P})}  \big]$ where the assumption $\Delta \mathbb{E}^{(t,\bsy{P})} > 0$ guarantees that the latter is strictly higher than the former. Consequently, if $j$ corresponds to the best correct atom, then the second concentration inequality is used since $\mathbb{E}\lbrack f_{j}^{(t, \bsy{P})} \rbrack > \alpha$. Conversely, the first concentration inequality will be used for incorrect atoms. We now consider the convex combination 
\begin{align*}
\alpha & = \lambda \mathbb{E}\big[ f_{j_c^{(t,\bsy{P})}}^{(t, \bsy{P})}  \big] + (1-\lambda) \max_{j \in \overline{\mathcal{S}}} \mathbb{E}\big[ f_{j}^{(t, \bsy{P})} \big]\\
& = (1-\overline{\lambda}) \mathbb{E}\big[ f_{j_c^{(t,\bsy{P})}}^{(t, \bsy{P})} \big] + \overline{\lambda} \max_{j \in \overline{\mathcal{S}}} \mathbb{E}\big[ f_{j}^{(t, \bsy{P})} \big]
\end{align*}
where $\overline{\lambda} = 1 - \lambda$. Thus, we obtain
\begin{align*}
\mathbb{P} \left(f_{j_c^{(t,\bsy{P})}}^{(t, \bsy{P})} \leq  \alpha \right) & \leq \exp \left( - \dfrac{\left(\mathbb{E}\lbrack f_{j_c^{(t,\bsy{P})}}^{(t, \bsy{P})} \rbrack - \alpha  \right)^2}{2 \| \bsy{\sigma}_{j_c^{(t,\bsy{P})}}^{(\bsy{P})} \|_2^2} \right) \\
& = \exp \left( - \dfrac{\overline{\lambda}^2}{2 \| \bsy{\sigma}_{j_c^{(t,\bsy{P})}}^{(\bsy{P})} \|_2^2} \left(\Delta \mathbb{E}^{(t,\bsy{P})}\right)^2 \right).
\end{align*}
For $j \in \overline{\mathcal{S}}$, we obtain
\begin{align*}
\hspace*{-2mm}
\mathbb{P} \left(f_{j}^{(t, \bsy{P})} \geq   \alpha \right) & \leq \exp \left( - \dfrac{ \left(\alpha - \mathbb{E}\lbrack f_{j}^{(t, \bsy{P})} \rbrack \right)^2}{2 \| \bsy{\sigma}_{j}^{(\bsy{P})} \|_2^2} \right) \\
& \leq \exp \left( - \dfrac{\left( \alpha - \max_{\tilde{j} \in \overline{\mathcal{S}}} \mathbb{E}\lbrack f_{\tilde{j}}^{(t, \bsy{P})} \rbrack \right)^2}{2 \| \bsy{\sigma}_{j}^{(\bsy{P})} \|_2^2} \right) \\
& = \exp \left( - \dfrac{\lambda^2}{2 \| \bsy{\sigma}_{j}^{(\bsy{P})} \|_2^2} \left(\Delta \mathbb{E}^{(t,\bsy{P})}\right)^2 \right)
\end{align*}
where the second inequality holds because we have $\alpha > \max_{\tilde{j} \in \overline{\mathcal{S}}} \mathbb{E}\lbrack f_{\tilde{j}}^{(t, \bsy{P})} \rbrack \geq \mathbb{E}\lbrack f_{j}^{(t, \bsy{P})} \rbrack$ for every $j \in \overline{\mathcal{S}}$. A less sharp upper bound is obtained for both inequalities by remembering that $\| \bsy{\sigma}_{j}^{(\bsy{P})} \|_2 \leq \| \bsy{\sigma} \|_2$. Setting $\lambda = \overline{\lambda} = 0.5$ and combining Theorem~\ref{thm:basicProbIneq} with the derived inequalities for $j = j_c^{(t,\bsy{P})}$ and $j \in \overline{\mathcal{S}}$ conclude the proof.

\subsection{Proof of Theorem~\ref{thm:finalThmNew}}

The event $E_{\mathrm{succ}}^{(t, \bsy{P})}$ occurs whenever SOMP picks a correct atom at iteration $t$ given the projection matrix $\bsy{P} \in \mathcal{P}^{(t)}$. Considering all the possible orthogonal projectors $\bsy{P} \in \mathcal{P}^{(t)}$ from iteration $0$ to iteration $s$ included, we have $\mathcal{C}_s = \sum_{t=0}^{s} \binom{|\mathcal{S}|}{t}$ possible orthogonal projectors. If SOMP succeeds in choosing a correct atom for all the possible orthogonal projectors at each iteration, then we know that correct decisions will occur at each iteration. Thus, defining $E_{\mathrm{succ}}^{(s)}$ as the event occurring whenever SOMP is successful during the first $s+1$ iterations, we have 
\begin{equation*}
\mathbb{P} \lbrack E_{\mathrm{succ}}^{(s)} \rbrack \geq \mathbb{P} \left\lbrack \bigcap_{t=0}^s \bigcap_{\bsy{P} \in \mathcal{P}^{(t)}} E_{\mathrm{succ}}^{(t, \bsy{P})} \right\rbrack  = 1 - \mathbb{P} \left\lbrack \bigcup_{t=0}^s \bigcup_{\bsy{P} \in \mathcal{P}^{(t)}} E_{\mathrm{fail}}^{(t, \bsy{P})} \right\rbrack
\end{equation*}
where the event intervening in the R.H.S. is the complementary event of that of the preceding expression. In particular, $ E_{\mathrm{fail}}^{(t, \bsy{P})}$ is the event occurring when SOMP picks an incorrect atom at iteration $t$ given the orthogonal projector $\bsy{P} \in \mathcal{P}^{(t)}$.  The union bound yields 
\begin{equation*}
\mathbb{P} \left\lbrack \bigcup_{t=0}^s \bigcup_{\bsy{P} \in \mathcal{P}^{(t)}} E_{\mathrm{fail}}^{(t, \bsy{P})} \right\rbrack \leq \sum_{t=0}^s \sum_{\bsy{P} \in \mathcal{P}^{(t)}} \mathbb{P} \lbrack E_{\mathrm{fail}}^{(t, \bsy{P})} \rbrack.
\end{equation*}
Therefore, if $E_{\mathrm{fail}}^{(s)}$ is the probability of failure of SOMP during the first $s+1$ iterations, we have $\mathbb{P} \lbrack E_{\mathrm{fail}}^{(s)} \rbrack = 1 - \mathbb{P} \lbrack E_{\mathrm{succ}}^{(s)} \rbrack \leq \sum_{t=0}^s \sum_{\bsy{P} \in \mathcal{P}^{(t)}} \mathbb{P} \lbrack E_{\mathrm{fail}}^{(t, \bsy{P})} \rbrack$ where Theorem~\ref{thm:concentratedProbIneq} yields
\begin{equation*}
\mathbb{P} \left[ E_{\mathrm{fail}}^{(t, \bsy{P})} \right] \leq (n - |\mathcal{S}| + 1) \exp \left[ - \dfrac{1}{8 \| \bsy{\sigma} \|_2^2} (\Delta \mathbb{E}^{(t,\bsy{P})})^2 \right].
\end{equation*}
The lower bound for $\Delta \mathbb{E}^{(t,\bsy{P})}$ derived in Section~\ref{subsec:lowBoundDeltaE} provides 
\begin{equation*}
\Delta \mathbb{E}^{(t,\bsy{P})} \geq \left(1 - \frac{1}{\Gamma} \right) \psi  \tau_X - \sqrt{\frac{2}{\pi}} \| \bsy{\sigma} \|_1.
\end{equation*}
Using the inequality $n - |\mathcal{S}| + 1 \leq n$, noticing that the upper bound of $\mathbb{P} \lbrack E_{\mathrm{fail}}^{(s)} \rbrack$ has become independent of $t$ and $\bsy{P}$ and remembering that $|\mathcal{P}^{(t)}| = \binom{|\mathcal{S}|}{t}$ conclude the first part of the proof. Regarding the inequality on $\mathcal{C}_s$, we have \cite[Appendix A]{jacques2013robust}, $\sum_{t=1}^s \binom{|\mathcal{S}|}{t} \leq \binom{|\mathcal{S}|+s-1}{s}$ as well as $\binom{|\mathcal{S}|}{t} \leq \left( \frac{e |\mathcal{S}|}{t} \right)^t$. Thus, we easily obtain 
\begin{equation*}
\mathcal{C}_s = 1 + \sum_{t=1}^s \binom{|\mathcal{S}|}{t} \leq 1 + \left( \frac{e (|\mathcal{S}|+s-1)}{s} \right)^s.
\end{equation*}

\newpage
\nocite{*}
\bibliographystyle{IEEEtranS}
\bibliography{mybib}

\end{document}